\documentclass[english,aps,prl,amsmath,amssymb,twocolumn,letterpaper,showpacs,superscriptaddress]{revtex4-1}
\usepackage[T1]{fontenc}
\usepackage[latin9]{inputenc}
\usepackage{babel}
\usepackage{verbatim}

\usepackage[normalem]{ulem} 

\usepackage{chngcntr}
\counterwithin*{equation}{section} 
\counterwithin*{figure}{section} 

\usepackage{amsmath}
\usepackage{amssymb}
\usepackage{graphicx}
\usepackage{xcolor}
\usepackage[unicode=true,pdfusetitle,
 bookmarks=true,bookmarksnumbered=false,bookmarksopen=false,
 breaklinks=false,pdfborder={0 0 1},backref=false,colorlinks=false]
 {hyperref}

\makeatletter
\usepackage{graphicx}
\usepackage[T1]{fontenc}

\makeatother

\begin{document}
\title{Superconductivity of neutral modes in quantum Hall edges}
\author{Jukka I. V\"ayrynen}
\affiliation{Department of Physics and Astronomy, Purdue University, West Lafayette,
Indiana 47907 USA}
\author{Moshe Goldstein}
\affiliation{Raymond and Beverly Sackler School of Physics and Astronomy, Tel Aviv
University, Tel Aviv 6997801, Israel}
\author{Yuval Gefen}
\affiliation{Department of Condensed Matter Physics, Weizmann Institute of Science,
Rehovot 76100, Israel}
\date{\today}
\begin{abstract}
Edges of quantum Hall phases give rise to a multitude of exotic modes supporting quasiparticles of different values of charge and quantum statistics.  Among these are neutralons (chargeless anyons with semion statistics), which were found to be ubiquitous in fractional quantum Hall matter. Studying and manipulating the neutral sector is an intriguing and interesting challenge, all the more so since these particles are accessible experimentally. Here we address the limit of strongly-interacting neutralons giving rise to neutralon superconductivity, where pairing is replaced by a quarteting mechanism.  We discuss several manifestations of this effect, realizable in existing experimental platforms.  Furthermore, this superconducting gapping mechanism may be exploited to facilitate the observation of interference of the accompanying charged anyons. 
\end{abstract}
\maketitle
\emph{Introduction.} 
A two dimensional electron gas in a fractional
quantum Hall (FQH) state can host exotic anyonic quasiparticles and boundary modes~\citep{LeinaasMyrheim,PhysRevLett.49.957,PhysRevLett.50.1395,PhysRevLett.52.1583,PhysRevLett.53.722}. The boundary modes may be used as building blocks for realizing anyonic transport and designing interference experiments to observe fractional quantum statistics beyond  bosons and fermions~\cite{Feldman_2021}. 
Major recent experimental developments involve the observation~\citep{Bartolomei_2020} of Hong-Ou-Mandel~\citep{PhysRevLett.86.4628,PhysRevLett.109.106802,PhysRevB.88.235415,PhysRevLett.116.156802} anyonic correlations as well as a demonstration~\cite{nakamura2020direct} of anyonic interferometry~\citep{PhysRevB.55.2331,PhysRevLett.96.016802,PhysRevLett.96.016803,PhysRevLett.96.226803,PhysRevB.74.045319,PhysRevLett.97.216404,PhysRevB.76.085333,PhysRevLett.124.106805}.

Some of the exotic boundary modes arise as renormalized bare edge modes. Examples include neutral modes 
 which  have been
experimentally detected through thermometry~\cite{2012NatPh...8..676V} and  the generation of upstream charge
noise~\cite{2010Natur.466..585B,PhysRevLett.107.036805,PhysRevLett.108.226801,2012NatCo...3E1289G,2014NatCo...5E4067I}. 
The paradigmatic model of gapless neutral modes was introduced by
Kane, Fisher, and Polchinski (KFP)~\citep{kane_randomness_1994}
for the $\nu=2/3$ FQH edge which hosts counterpropagating $\nu=1$
and $\nu=1/3$ chiral bare modes~\citep{1990PhRvL..64..220M,PhysRevLett.64.2206}.
Random backscattering and Coulomb interaction between the modes drive the edge to a new low-energy fixed point that hosts a charge-$2e/3$ mode and a counterpropagating ``upstream'' neutral mode. 
In the original KFP model the neutral modes satisfy SU(2) symmetry, but more complex edges (for example, due to edge reconstruction) may give rise to more elaborate structures such as SU(3) symmetric  modes~\citep{wang_edge_2013}.

So far these studies have focused on non-interacting neutral modes. 
Exploring the physics of  interacting neutral mode system opens the door to new exotic phases.
Here, we show that interaction within the  neutral mode sector may give rise to ``\textit{neutralon} \textit{superconductivity}'', relying on amalgamating (hereafter ``pairing'') together a quartet of  neutral quasiparticles.

The neutral modes are chiral, so in order to open a superconducting gap,  a counterpropagating partner
needs to be introduced~\citep{wen_quantum_2004}. 
Here, we 
take advantage of a recent material engineering breakthrough~\citep{2018NatPh..14..411R}
and theoretically investigate a suitably designed FQH bilayer, where
two counterpropagating copies of the $\nu=2/3$ FQH neutral mode appear,
see also~\citep{wang2021transport}. 
We consider the limit of weak  neutralon-neutralon interactions.  
In that limit, 
in the presence of disorder-induced tunneling between the counterpropagating neutralons, 
Anderson localization is suppressed~\footnote{ We note that neutralons are semions, hence the allowed  backscattering process involves four neutralons, making it highly 	
	irrelevant at low energies~\citep{kane_randomness_1994,Kane_1997}. 
	(The scaling dimension of the neutralon backscattering operator is 2.)}, and hence will not compete against opening a superconducting gap. 
Uniform 
backscattering competes with uniform pairing, both being marginal operators (for weak interactions). 
However, neutralons are charge dipoles, hence subject to a weak attractive density-density interaction which favors  pairing. 
Interestingly, due to the semionic nature of neutralons,  the pairing must involve four of them. 
This pairing conserves momentum and is marginally relevant even in
the presence of edge disorder. 

The novel type of superconductivity in the neutral sector has experimental manifestations that involve measurement of the charge modes. 
Tunneling of electrons across a quantum point contact bridging fractional quantum Hall states will generally excite neutral modes. When the neutral modes are gapped by pairing, electron tunneling is highly suppressed at low energies, which may be observed in the low-bias $I-V$ characteristics and the shot noise Fano factor.  
Another signature  involves a confined quantum dot or anti-dot geometry, where unpaired neutralons come at a cost of pairing energy,  which is manifest in Coulomb blockade peak spacings~\citep{PhysRevLett.114.156401}.

Furthermore,
 our analysis concerns the design of anyonic interferometers.  
It is known that gapless neutralons may act as ``which-path'' detectors. 
Their ubiquity~\citep{2014NatCo...5E4067I,PhysRevLett.122.246801}
 leads to dephasing, hence suppression of   interference~\citep{PhysRevLett.117.276804}. 
We therefore %
propose that when neutralons condense to a gapped state, the sensitivity of anyonic interferometry will improve. 
Below, we outline interferometer designs that could be used to gap out the harmful neutralons while leaving the charge excitations gapless, thus improving interferometer performance. 
The observation of superconducting neutralon phase may have far-reaching impact on  the quest of anyonic interference.

Our theoretical analysis follows these steps: We will consider a bilayer of counterpropagating neutral modes. The latter are represented by bosonic fields as described in the original work by KFP~\citep{kane_randomness_1994}. The corresponding action, Eq.~(\ref{eq:SKFPFP}), is derived without inter-layer interactions. We then include weak inter-layer perturbations, pairing and backscattering, see Eq.~(\ref{eq:H_pairing1}). 
 Employing the perturbative renormalization group,  we identify the parameter regime where pairing becomes the most relevant perturbation. 
The energy scale at which such pairing becomes non-perturbative (strongly coupled) is the superconducting gap $\Delta_n$, at energies below which  the  pairing interaction leads to a \textit{quartet superconductivity}. 
Finally, we discuss the experimental manifestations of the superconducting phase.

\begin{figure}
\includegraphics[width=0.95\columnwidth]{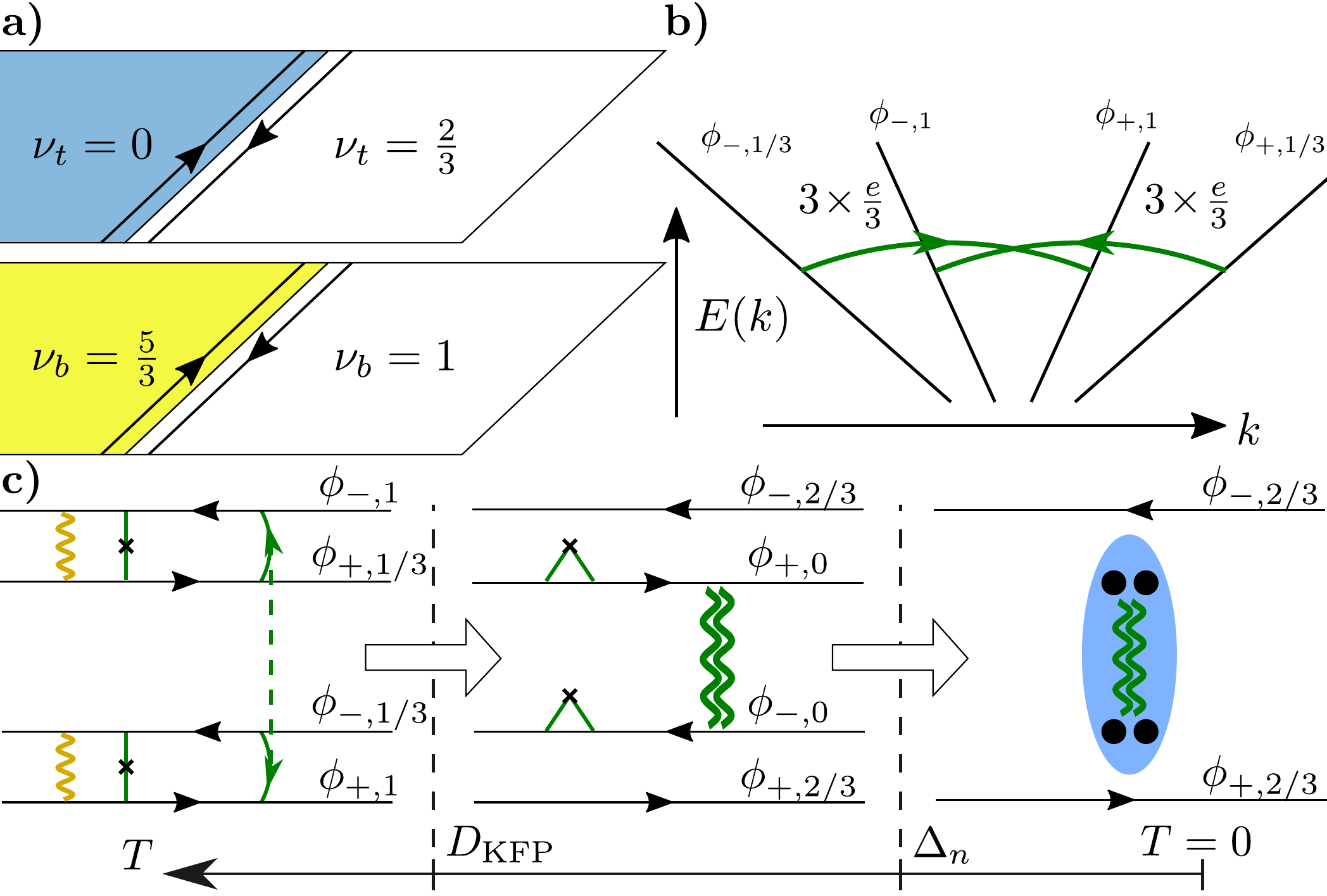}

\caption{\textbf{a)} Bilayer FQH structure where neutralons can be gapped.
We consider an interface where both bottom and top layers of the structure
have an interface described by the KFP theory for the $\nu=2/3$ FQH
state. The filling fractions are chosen in a way that produces opposite
spin polarizations for the top and bottom layers. \textbf{b)} Edge
spectrum of the clean limit. Correlated intralayer backscattering
process (green curved arrows) that contributes to pairing of neutralons
at low energies. \textbf{c)} The characteristic temperature scales
and the two-step RG flow towards low temperature (large white arrows).
The bare edge modes at high temperatures on the left, with Coulomb
interaction (yellow wiggly line), random backscattering (solid line
with a cross), and correlated intralayer backscattering (curved arrows connected by a dashed line). 
As temperature is lowered below $D_{\text{KFP}}$, the edge is described
by the KFP low-energy theory (middle panel) of $2e/3$ charge modes
and disordered neutral modes with a pairing interaction (green wiggly double line
line). At temperatures below $\Delta_{n}$, quartets of neutralons become
gapped (blue ellipse) and only the charge modes remain (right panel). Their opposite
spin polarization prevents backscattering. \label{fig:SingleEdge}}
\end{figure}

\begin{figure}
\includegraphics[width=0.95\columnwidth]{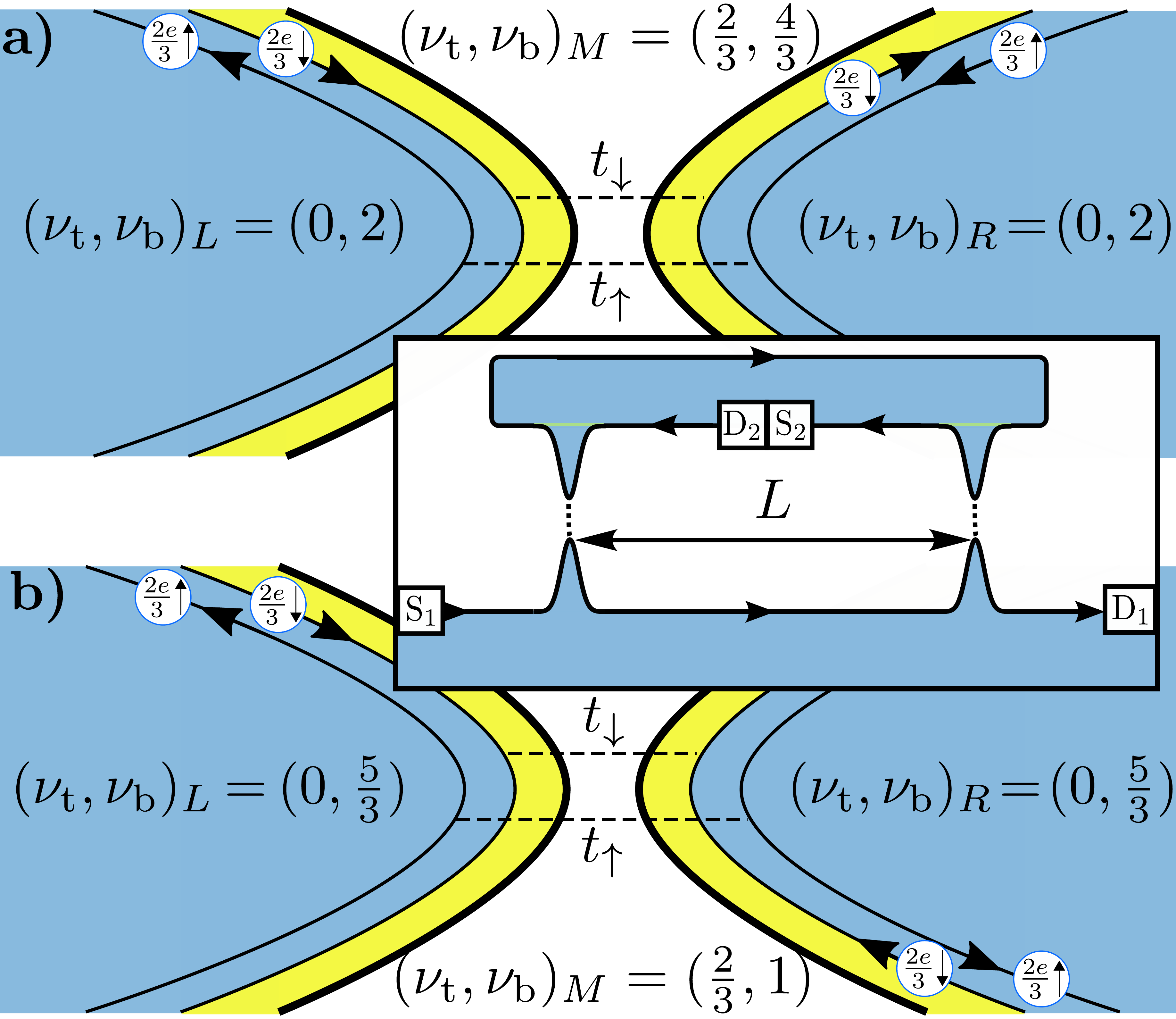}

\caption{Two quantum point contact (QPC) designs showing  the composite edge
of Fig.~\ref{fig:SingleEdge}a from the top. For the sake of clarity,
we have shifted the top and bottom layers laterally. Both sides of
the QPC are at the fixed point where neutral modes are localized.
Each side therefore hosts a helical pair of $2e/3$ charge modes with
spin up (down) mode living on the bottom (top) layer of the double
quantum well. Tunneling across the QPC is assumed to conserve the
spin eigenvalue. The figures show two different setups: in \textbf{a)}
the top layer quasiparticles tunnel through a $\nu=2/3$ bulk rather
than a trivial vacuum (in the bottom layer tunneling is through a
$\nu=1$ trivial vacuum). In \textbf{b)}, in both layers quasiparticles
tunnel through a fractional $\nu=2/3$ vacuum. These two setups have
different zero-bias anomalies in the tunneling current and different
shot noise Fano factors. Inset: A Mach-Zehnder interferometer is not
susceptible to dephasing from neutral modes provided its linear size
$L$ exceeds the neutral mode decay length, $L\gg v_{0}/\Delta_{n}$,
and the bias voltage is low, $eV\ll\Delta_{n}$. \label{fig:QPC} }
\end{figure}

\emph{Model of a single edge.} We start from the description
of a single composite interface depicted in Fig.~\ref{fig:SingleEdge}a.
Both the top and bottom components have an interface similar to a
$\nu=2/3$ edge. Assuming spin-polarized Landau levels, our edge theory
therefore consists of four chiral bosonic fields $\phi_{+,1/3,\uparrow},\,\phi_{-,1,\uparrow},\,\phi_{-,1/3,\downarrow},\,\phi_{+,1,\downarrow}$
where the subscripts indicate the chirality (``$+$'' denotes a right
mover), charge, and spin. The imaginary time action is~\citep{kane_randomness_1994}
($a$ is a short-distance cutoff)
\begin{flalign}
S^{(0)} & =\int d\tau dx\frac{1}{4\pi}\left[\partial_{x}\boldsymbol{\Phi}\mathbf{K}i\partial_{\tau}\boldsymbol{\Phi}+\partial_{x}\boldsymbol{\Phi}\mathbf{V}\partial_{x}\boldsymbol{\Phi}\right]\label{eq:ShomogDiab}\\
 & +\frac{1}{a}\int d\tau dx\sum_{l=\mathrm{t,b}}\left[\xi_{l}(x)e^{i\mathbf{c}_{l}\cdot\boldsymbol{\phi}_{l}}+\xi_{l}^{*}(x)e^{-i\mathbf{c}_{l}\cdot\boldsymbol{\phi}_{l}}\right]\,,\nonumber 
\end{flalign}
where we introduced the 4-component vector $\boldsymbol{\Phi}=(\boldsymbol{\phi}_{\mathrm{t}},\boldsymbol{\phi}_{\mathrm{b}})=(\phi_{+,1/3,\uparrow},\phi_{-,1,\uparrow},\phi_{-,1/3,\downarrow},\phi_{+,1,\downarrow})$.
In this basis the matrix $\mathbf{V}$ is almost block-diagonal~\footnote{See Supplemental Material, which includes Refs.~\citep{PhysRevB.84.195436,PhysRevB.36.4581}.},
describing the velocities and short-range screened Coulomb interactions
between the modes; the matrix $\mathbf{K}=\text{diag}(3,-1,-3,1)$
describes the commutation relations~\footnote{The chiral fields obey $[\phi_{i}(x),\phi_{j}(x')]=\pi i(\mathbf{K}^{-1})_{ij}\text{sgn}(x-x')$~\citep{2017AnPhy.385..287P}.
}. We may use the same action to describe other interfaces such as
those depicted in Fig.~\ref{fig:QPC}~\cite{Note2}. On the second
line of Eq.~(\ref{eq:ShomogDiab}) we have included random intralayer
backscattering of electrons between the counterpropagating $1/3$
and $1$ modes; here $\mathbf{c}_{t(b)}=(-1)(3,1)$ and $\xi_{l}$
is a $\delta$-correlated random coefficient, $\left\langle \xi_{l}(x)\xi_{l}^{*}(x')\right\rangle =a^{-1}W_{l}\delta(x-x')$,
with zero average. This term is a relevant perturbation under the
renormalization group~\citep{giamarchi_anderson_1988} (RG) and leads
to a non-trivial renormalization of the edge theory~\citep{kane_randomness_1994}.
In Eq.~(\ref{eq:ShomogDiab}) we neglect inter-layer perturbations
which will be included later, see Eq.~(\ref{eq:H_pairing1}). 

Below a temperature scale $T\sim D_{\text{KFP}}$, the disorder
strength $W_{l}$ becomes large~\footnote{We may estimate $D_{\text{KFP}}$ by using the RG equation $dW/d\ln D^{-1}=(3-2\delta)W$
where $\delta<3/2$ is the scaling dimension and $D$ the reduced
bandwidth. At strong coupling we have $W(D_{\text{KFP}})\sim v_{0}^{2}$
which yields $D_{\text{KFP}}\sim D_{0}(W(D_{0})/D_{0})^{1/(3-2\delta)}$
in terms of the bare bandwidth $D_{0}$. 
If the KFP fixed point is not fully
reached (say, at temperature $T\gtrsim D_{\text{KFP}}$) there will
be RG irrelevant interactions such as $\partial_{x}\phi_{+,2/3,\downarrow}\partial_{x}\phi_{-,0}$
between the charge and neutral modes. }
 and the edge action can be diagonalized in terms of spinless neutralons
 and spinful charge-$2e/3$ modes given by the respective linear
combinations (Fig.~\ref{fig:SingleEdge}a),  
\begin{equation}
\phi_{+,0}\!=\!\frac{3\phi_{+,1/3,\uparrow}+\phi_{-,1,\uparrow}}{\sqrt{2}},\,\phi_{-,2/3,\uparrow}\!=\!\sqrt{\frac{3}{2}}[\phi_{+,1/3,\uparrow}+\phi_{-,1,\uparrow}],\label{eq:neutralcharge}
\end{equation}
and similarly for the bottom edge. Introducing $\boldsymbol{\phi}=(\phi_{+,2/3,\downarrow},\phi_{-,2/3,\uparrow},\phi_{+,0},\phi_{-,0})$,
the low-energy action is 
\begin{flalign}
 & S_{\text{KFP}}^{(0)}=\int d\tau dx\frac{1}{4\pi}\left[\partial_{x}\boldsymbol{\phi}\mathbf{K}_{\text{KFP}}i\partial_{\tau}\boldsymbol{\phi}+\partial_{x}\boldsymbol{\phi}\mathbf{V}_{\text{KFP}}\partial_{x}\boldsymbol{\phi}\right]\label{eq:SKFPFP}\\
 & +\frac{1}{a}\int d\tau dx\left[\xi_{t}(x)e^{i\sqrt{2}\phi_{+,0}}+\xi_{b}(x)e^{-i\sqrt{2}\phi_{-,0}}\!+\!\text{h.c.}\right],\nonumber 
\end{flalign}
with $\mathbf{K}_{\text{KFP}}=\text{diag}(1,-1,1,-1)$ and $\mathbf{V}_{\text{KFP}}$
is a block diagonal matrix. The block diagonality of $\mathbf{V}_{\text{KFP}}$
is a result of the random intralayer backscattering, which makes neutralon-chargon
interactions $\partial_{x}\phi_{\pm,0}\partial_{x}\phi_{\pm,2/3}$
irrelevant~\citep{kane_randomness_1994}. However, $\mathbf{V}_{\text{KFP}}$
includes a neutralon-neutralon interaction $v_{0,0}$ which is not
irrelevant for layer-correlated disorder (considered below)~\cite{Note2}.
The second line in Eq.~(\ref{eq:SKFPFP}) introduces random phases
into the neutral sector but does not give rise to a gap~\citep{kane_randomness_1994}. 

\emph{Inter-layer tunneling. }Let us next include weak inter-layer
tunneling to the action $S_{\text{KFP}}^{(0)}$, Eq.~(\ref{eq:SKFPFP}).
This introduces the leading (in the RG sense) perturbations in the
neutral sector: pairing {[}depicted in Fig.~\ref{fig:SingleEdge}b{]},
$O_{p}=e^{i\sqrt{2}[\phi_{+,0}-\phi_{-,0}]}$, and backscattering,
\emph{$O_{b}=e^{i\sqrt{2}[\phi_{+,0}+\phi_{-,0}]}$}. In the absence
of neutral-neutral interactions ($v_{0,0}=0$), both operators have
a scaling dimension $\delta=2$ and they are thus marginal (to leading
order) as homogeneous perturbations~\cite{Note1}. However, for $v_{0,0}\neq0$,
one of the two operators is favored: for negative (positive) $v_{0,0}$,
pairing (backscattering) becomes relevant while backscattering (pairing)
becomes irrelevant. The relevant pairing term gives rise to a gap
in the neutralon spectrum, $\Delta_{n}\approx D_{\text{KFP}}|\lambda_{p}|^{v_{0}/(2|v_{0,0}|)}$,
in the limit $\lambda_{p}\ll|v_{0,0}|/v_{0}\ll1$ where $\lambda_{p}$
is the dimensionless pairing amplitude~\cite{giamarchi_quantum_2003,Note2}.
We show below that in the case when $v_{0,0}/v_{0}$ is comparable
to the pairing and backscattering amplitudes, all three interactions get
significantly renormalized but the general conclusion of a gap remains. 
 The backscattering operator does not conserve
momentum (unlike pairing), and is expected to be less relevant
when the neutralons have a finite density (as depicted in Fig.~\ref{fig:SingleEdge}b).

 In the charge sector, backscattering is forbidden by spin conservation.
The pairing of the charge modes is highly irrelevant~\footnote{For a range of bare parameters, attraction may develop between the
charge modes, in which case pairing becomes relevant~\citep{Vayrynen_2019}.
Here we assume that this is not the case.} and also forbidden by charge conservation in the absence of an external
superconductor~\citep{PhysRevX.3.021009}.

Next we will analyze the interlayer pairing $O_{p}$ and backscattering
$O_{b}$ in the neutral sector. It is convenient to introduce the
$SU(2)_{1}$ current operators~\citep{kane_randomness_1994,2017AnPhy.385..287P}
\begin{equation}
J_{\tau}^{z}\!=\!\frac{1}{2\pi\sqrt{2}}\partial_{x}\phi_{\tau,0},\,\,J_{\tau}^{\pm}\!=\!\frac{1}{2\pi a}e^{\pm i\tau\sqrt{2}\phi_{\tau,0}},\,\tau=t/b=+/-,
\end{equation}
and $J_{\tau}^{\pm}=J_{\tau}^{x}\pm iJ_{\tau}^{y}$. In terms of the
currents, we have $O_{p}=2\pi aJ_{t}^{+}J_{b}^{+}$ and \emph{$O_{b}=2\pi aJ_{t}^{+}J_{b}^{-}$.
}We can write the combined neutralon inter-layer Hamiltonian in the
form 
\begin{flalign}
H_{p+b}\negmedspace & =\negmedspace2\pi v_{0}\negthinspace\int dx\negmedspace\sum_{i=x,y,z}\negmedspace\lambda^{i}J_{t}^{i}J_{b}^{i}\,,\label{eq:H_pairing1}
\end{flalign}
where  $\lambda^{x}=\lambda_{p}+\lambda_{b}$, $\lambda^{y}=\lambda_{b}-\lambda_{p}$
and $\lambda_{p},\,\lambda_{b}$ are the dimensionless pairing and
backscattering amplitudes and $v_{0}$ is the neutralon velocity.
The neutralon density-density interaction from Eq.~(\ref{eq:SKFPFP})
is included in the ZZ term, $\lambda_{0}^{z}=2v_{0,0}/v_{0}$. Upon
reducing the bandwidth, these coupling constants get renormalized.
 In the absence of disorder {[}the second line in Eq.~(\ref{eq:SKFPFP}){]},
the perturbative RG equations for $\lambda^{i=x,y,z}$ are~\citep{gogolin2004bosonization}:
\begin{flalign}
\frac{d}{dl}\lambda^{x} & =\lambda^{y}\lambda^{z}\,,\quad\frac{d}{dl}\lambda^{y}=\lambda^{x}\lambda^{z}\,,\label{eq:Rg1}\\
\frac{d}{dl}\lambda^{z} & =\lambda^{x}\lambda^{y}\,,\quad(l=\ln D_{\text{KFP}}/D)\,,\label{Rg2}
\end{flalign}
where $D\ll D_{\text{KFP}}$ is the reduced bandwidth. We solve the
above equations for $\lambda^{i}(D)$ with the initial condition $\boldsymbol{\lambda}(D_{\text{KFP}})=(\lambda_{p}+\lambda_{b},\lambda_{b}-\lambda_{p},\lambda_{0}^{z})^{T}$.
We assume that pairing and backscattering are weak, so that $|\lambda_{0}^{z}|>|\lambda^{x}|,|\lambda^{y}|$.
Then, the sign of $\lambda_{0}^{z}$ determines the low-energy RG
fixed point: when $\lambda_{0}^{z}>0$, the fixed point corresponds
to strong backscattering ($\lambda^{x}=\lambda^{y}=\pm\lambda^{z}$),
while if $\lambda_{0}^{z}<0$, the fixed point is of strong pairing
type ($\lambda^{x}=-\lambda^{y}=\pm\lambda^{z}$). Within each type,
the fixed point is further determined by the sign of $\lambda_{b}$
or $\lambda_{p}$: for example, in the strong pairing case $\lambda_{p}>0$
flows to ($\lambda^{x}=-\lambda^{y}=-\lambda^{z}>0$) while $\lambda_{p}<0$
flows to ($\lambda^{x}=-\lambda^{y}=+\lambda^{z}<0$). 

To estimate the strong coupling energy scale $\Delta_{n}$, we set
$|\lambda^{i}(\Delta_{n})|\gg1$. We find $\Delta_{n}\approx D_{\text{KFP}}\left(2\frac{|\lambda_{0}^{z}|}{|\lambda_{p}|}\right)^{-\frac{1}{|\lambda_{0}^{z}|}}$
in the limit $|\lambda_{p}|\ll|\lambda_{0}^{z}|\ll1$ and $\Delta_{n}\approx D_{\text{KFP}}e^{-\pi/2|\lambda_{p}|}$
in the limit $|\lambda_{0}^{z}|\ll|\lambda_{p}|\ll1$~\cite{Note2}. At temperatures $T\ll\Delta_{n}$, the neutral excitations are gapped
and only the charge modes remain from Eq.~(\ref{eq:SKFPFP}). 
Next, we will show that 
the random terms ${\propto\xi_{t}(x),}\,\xi_{b}(x)$
in Eq.~(\ref{eq:SKFPFP}) do not modify our conclusions. 

Interpreting the current operators $\mathbf{J}_{t,b}$ as spin densities,
the second line of Eq.~(\ref{eq:SKFPFP}) can be regarded as a random
``in-plane magnetic field''; the Hamiltonian corresponding to Eq.~(\ref{eq:SKFPFP})
reads 
\begin{equation}
H_{\text{neutral}}=2\pi\sum_{\tau = t,b}\int dx\left(\frac{1}{3}v_{0}\mathbf{J}_{\tau}^{2}+\xi_{\tau}(x)J_{\tau}^{+}+\xi_{\tau}^{*}(x)J_{\tau}^{-}\right) \! .\label{eq:HNeutral}
\end{equation}
 The random magnetic field can be cancelled by the following gauge
transformation, that preserves the $SU(2)_{1}$ algebra~\citep{2017AnPhy.385..287P}
for $\tau=t,b$, 
\begin{equation}
J_{\tau}^{i}=S_{\tau}^{ij}\tilde{J}_{\tau}^{j}+\frac{1}{8\pi}\varepsilon^{ijk}[S_{\tau}\partial_{x}S_{\tau}^{T}]^{jk}\,,\label{eq:JTransformation}
\end{equation}
 where $S_{\tau}(x)$ is a suitably chosen~\cite{Note2} real orthogonal
matrix. For generic disorder, Eq.~(\ref{eq:JTransformation}) does
not keep the pairing term invariant and finding the ground state configuration
is difficult.\emph{ }However, in the simple and realistic case of
layer-correlated disorder, $\xi_{t}=\xi_{b}^{*}\equiv\xi$, we have~\cite{Note2}
\begin{flalign}
\mathbf{J}_{t}^{T}\boldsymbol{\eta}\mathbf{J}_{b} & =\mathbf{\tilde{J}}_{t}^{T}\boldsymbol{\eta}\mathbf{\tilde{J}}_{b}\,,\quad\text{where }\boldsymbol{\eta}=\text{diag}(1,-1,-1)\,.\label{eq:PairingTransformation}
\end{flalign}
Thus, the rotation~(\ref{eq:JTransformation}) makes the Hamiltonian
independent of disorder, 
\begin{equation}
	 H_{\text{neutral}}+H_{\text{pairing}}=2\pi v_{0} \int \! dx \left[ \frac{1}{3}\sum_{\tau=t,b}\mathbf{\tilde{J}}_{\tau}^{2}
	 + \lambda\mathbf{\tilde{J}}_{t}^{T}\boldsymbol{\eta}\mathbf{\tilde{J}}_{b} \right] ,\label{eq:HneutralPairingRotated}
\end{equation}
as long as we have $\boldsymbol{\lambda}=(\lambda,-\lambda,-\lambda)^{T}$
in Eq.~(\ref{eq:H_pairing1}). We can therefore use Eqs.~(\ref{eq:Rg1})--(\ref{Rg2}),
derived in the absence of disorder, to study Eq.~(\ref{eq:HneutralPairingRotated}).
With $\lambda>0$, we find a strong-pairing RG fixed point which preserves
the direction of the vector $\boldsymbol{\lambda}$. 
We expect that the fixed point with $\lambda<0$ is similarly stable to disorder~\cite{Note2}.

We have shown that, under certain assumptions, the disorder term in
Eq.~(\ref{eq:HNeutral}) can be gauged away and the same low-energy
fixed points as in the clean system can be reached. When $\lambda_{0}^{z}<0$,
we identified two stable strong-pairing fixed points corresponding
to $\lambda^{x}>0$ and $\lambda^{x}<0$. Next, we study the low-energy
properties of the charge excitations near a fixed point where the
neutralons are paired. 

\emph{Experimental manifestations. } The gapping of
neutral modes at low energies has a number of implications to transport
experiments. Signatures of neutral mode gap can be found in tunneling
across a QPC, see Fig.~\ref{fig:QPC}. Tunneling of fractional charge
between the charge-$2e/3$ eigenmodes at low bias voltage may be impeded
in several ways depending on the filling factors of the left and
right sides of the QPC as well as the filling \emph{$(\nu_{\mathrm{t}},\nu_{\mathrm{b}})_{M}$}
of the middle section. Most conducive to fractional charge tunneling
is having fractional \emph{$(\nu_{\mathrm{t}},\nu_{\mathrm{b}})_{M}$}
(see Fig.~\ref{fig:QPC}a); in that case tunneling of $2e/3$ and
$e/3$ quasiparticles is allowed. 
The latter involves the gapped neutralons and is thus suppressed (similarly to the case of charge-$e$ tunneling discussed below) but the former is not. Indeed, 
the tunneling operator
$O_{2/3,\uparrow}=e^{-i\sqrt{\frac{2}{3}}\phi_{-,2/3,\uparrow,L}}e^{i\sqrt{\frac{2}{3}}\phi_{-,2/3,\uparrow,R}}$ 
creates (annihilates) a charge-$2e/3$ eigenmode on the right (left)
side of the QPC. The scaling dimension of $O_{2/3,\uparrow}$ is $\delta=2/3$
and the tunneling current shows the corresponding zero-bias anomaly,
$I\propto V^{2\delta-1}$ (keeping $eV\gg T$). The fractional tunneling
charge also has a noise signature~\citep{PhysRevLett.72.724,PhysRevLett.86.4628,PhysRevLett.103.236802}:
tunneling charges $2e/3$ leads to a shot noise Fano factor $2/3$. 

Tunneling is much more restricted when the middle region consists
of an integer filling fraction state, c.f. $\nu_{b}$ in the middle
section of Fig.~\ref{fig:QPC}b. In this case, only electrons (charge-$e$)
are allowed to tunnel through the middle section. However, tunneling
single electrons would excite the neutral modes and therefore come
at a high energy cost of order~\citep{PhysRevLett.114.156401} $\Delta_{n}$
(for voltage bias $eV\ll\Delta_{n}$). Tunneling of a pair of electrons (3 charge-$2e/3$ quasiparticles) does not excite the neutrals and is allowed.  (Also, tunneling of a ``Cooper pair'' of counterpropagating neutralons would be allowed but will not transfer charge.)   
   Tunneling of a pair of electrons  has a scaling dimension $\delta=4$,
suppressing the tunneling current at low bias, $I\propto V^{7}$.
 Thus, when the tunnel barrier (in either bottom or top layer) has
an integer filling fraction state, the low-bias tunneling current
becomes highly suppressed. The tunneling current shot noise Fano factor
in this case is expected to be $2$, yet its observation may be challenging
due to the smallness of the current. Gapped neutralons cannot propagate
along the edge and thus are not expected to produce noise. 

A complementary signature of neutralon pairing can be found in Coulomb
blockaded quantum dots or antidots~\citep{PhysRevLett.114.156401}.
In~\cite{Note2} we show that neutralon pairing gap leads to
a unique signature in the Coulomb blockade peak spacings.

The bilayer geometry where the neutral modes become gapped allows
one to consider anyonic Mach-Zehnder or Fabry-Perot interferometers
free of neutral mode dephasing, c.f. Fig.~\ref{fig:QPC}. As discussed
above, the configurations with fractional filling factors $(\nu_{\mathrm{t}},\nu_{\mathrm{b}})_{M}$
depicted in Fig.~\ref{fig:QPC}a are most suitable for constructing
such an interferometer since they allow tunneling of fractional charge
quasiparticles. The size of the neutral mode gap $\Delta_{n}$ imposes
some limitations to the interferometer design. For example, the distance
$L$ between the QPCs should be large enough, $L\gg v_{0}/\Delta_{n}$,
and the bias voltage low enough, $eV\ll\Delta_{n}$, so that neutral
modes cannot propagate through the interferometer causing dephasing. 
To observe interference, the length of the edge should not exceed the
full incoherence length scale~\citep{2017AnPhy.385..287P}. 

\emph{Discussion.} 
We showed that counterpropagating neutral modes %
 in a suitably designed FQH 
interface can be renormalized to a new type of superconducting phase with a neutralon quartet pairing. 
We focused on engineered bilayer
interfaces whose edge structure is similar to the $\nu=2/3$ KFP edge
theory~\citep{kane_randomness_1994}. 
In this case, the neutralons are semions and the superconductivity arises from neutralon quarteting. 
We expect our
 mechanism to also apply to reconstructed edges with
emergent chiral modes~\citep{wang_edge_2013} and other filling fractions, 
as long as these edges come with counterpropagating  neutral modes. 
With different types of edge structures other unconventional neutralon statistics may arise, and we anticipate even more exotic (superconducting) phases of strongly-interacting neutralon matter.

\begin{acknowledgments}
\emph{Acknowledgements.} We thank Jinhong Park for useful
discussions.  M.G. was supported by the Israel Science Foundation
(Grant No. 227/15) and the US-Israel Binational Science Foundation
(Grant No. 2016224). Y.G. was supported by DFG RO 2247/11-1, MI 658/10-2 and  CRC
183 (project C01), the Minerva Foundation, the German Israeli
Foundation (GIF I-1505-303.10/2019), the Helmholtz International Fellow Award, and by the Italia-Israel QUANTRA grant. 
\end{acknowledgments}

\bibliographystyle{apsrev4-1}
\bibliography{refs}

\providecommand{\noopsort}[1]{}\providecommand{\singleletter}[1]{#1}%
\begin{thebibliography}{52}%
\makeatletter
\providecommand \@ifxundefined [1]{%
 \@ifx{#1\undefined}
}%
\providecommand \@ifnum [1]{%
 \ifnum #1\expandafter \@firstoftwo
 \else \expandafter \@secondoftwo
 \fi
}%
\providecommand \@ifx [1]{%
 \ifx #1\expandafter \@firstoftwo
 \else \expandafter \@secondoftwo
 \fi
}%
\providecommand \natexlab [1]{#1}%
\providecommand \enquote  [1]{``#1''}%
\providecommand \bibnamefont  [1]{#1}%
\providecommand \bibfnamefont [1]{#1}%
\providecommand \citenamefont [1]{#1}%
\providecommand \href@noop [0]{\@secondoftwo}%
\providecommand \href [0]{\begingroup \@sanitize@url \@href}%
\providecommand \@href[1]{\@@startlink{#1}\@@href}%
\providecommand \@@href[1]{\endgroup#1\@@endlink}%
\providecommand \@sanitize@url [0]{\catcode `\\12\catcode `\$12\catcode
  `\&12\catcode `\#12\catcode `\^12\catcode `\_12\catcode `\%12\relax}%
\providecommand \@@startlink[1]{}%
\providecommand \@@endlink[0]{}%
\providecommand \url  [0]{\begingroup\@sanitize@url \@url }%
\providecommand \@url [1]{\endgroup\@href {#1}{\urlprefix }}%
\providecommand \urlprefix  [0]{URL }%
\providecommand \Eprint [0]{\href }%
\providecommand \doibase [0]{http://dx.doi.org/}%
\providecommand \selectlanguage [0]{\@gobble}%
\providecommand \bibinfo  [0]{\@secondoftwo}%
\providecommand \bibfield  [0]{\@secondoftwo}%
\providecommand \translation [1]{[#1]}%
\providecommand \BibitemOpen [0]{}%
\providecommand \bibitemStop [0]{}%
\providecommand \bibitemNoStop [0]{.\EOS\space}%
\providecommand \EOS [0]{\spacefactor3000\relax}%
\providecommand \BibitemShut  [1]{\csname bibitem#1\endcsname}%
\let\auto@bib@innerbib\@empty
\bibitem [{\citenamefont {Leinaas}\ and\ \citenamefont
  {Myrheim}(1977)}]{LeinaasMyrheim}%
  \BibitemOpen
  \bibfield  {author} {\bibinfo {author} {\bibfnamefont {J.}~\bibnamefont
  {Leinaas}}\ and\ \bibinfo {author} {\bibfnamefont {J.}~\bibnamefont
  {Myrheim}},\ }\href {\doibase 10.1007/BF02727953} {\bibfield  {journal}
  {\bibinfo  {journal} {Nuovo Cim. B}\ }\textbf {\bibinfo {volume} {37}},\
  \bibinfo {pages} {1} (\bibinfo {year} {1977})}\BibitemShut {NoStop}%
\bibitem [{\citenamefont {Wilczek}(1982)}]{PhysRevLett.49.957}%
  \BibitemOpen
  \bibfield  {author} {\bibinfo {author} {\bibfnamefont {F.}~\bibnamefont
  {Wilczek}},\ }\href {\doibase 10.1103/PhysRevLett.49.957} {\bibfield
  {journal} {\bibinfo  {journal} {Phys. Rev. Lett.}\ }\textbf {\bibinfo
  {volume} {49}},\ \bibinfo {pages} {957} (\bibinfo {year} {1982})}\BibitemShut
  {NoStop}%
\bibitem [{\citenamefont {Laughlin}(1983)}]{PhysRevLett.50.1395}%
  \BibitemOpen
  \bibfield  {author} {\bibinfo {author} {\bibfnamefont {R.~B.}\ \bibnamefont
  {Laughlin}},\ }\href {\doibase 10.1103/PhysRevLett.50.1395} {\bibfield
  {journal} {\bibinfo  {journal} {Phys. Rev. Lett.}\ }\textbf {\bibinfo
  {volume} {50}},\ \bibinfo {pages} {1395} (\bibinfo {year}
  {1983})}\BibitemShut {NoStop}%
\bibitem [{\citenamefont {Halperin}(1984)}]{PhysRevLett.52.1583}%
  \BibitemOpen
  \bibfield  {author} {\bibinfo {author} {\bibfnamefont {B.~I.}\ \bibnamefont
  {Halperin}},\ }\href {\doibase 10.1103/PhysRevLett.52.1583} {\bibfield
  {journal} {\bibinfo  {journal} {Phys. Rev. Lett.}\ }\textbf {\bibinfo
  {volume} {52}},\ \bibinfo {pages} {1583} (\bibinfo {year}
  {1984})}\BibitemShut {NoStop}%
\bibitem [{\citenamefont {Arovas}\ \emph {et~al.}(1984)\citenamefont {Arovas},
  \citenamefont {Schrieffer},\ and\ \citenamefont
  {Wilczek}}]{PhysRevLett.53.722}%
  \BibitemOpen
  \bibfield  {author} {\bibinfo {author} {\bibfnamefont {D.}~\bibnamefont
  {Arovas}}, \bibinfo {author} {\bibfnamefont {J.~R.}\ \bibnamefont
  {Schrieffer}}, \ and\ \bibinfo {author} {\bibfnamefont {F.}~\bibnamefont
  {Wilczek}},\ }\href {\doibase 10.1103/PhysRevLett.53.722} {\bibfield
  {journal} {\bibinfo  {journal} {Phys. Rev. Lett.}\ }\textbf {\bibinfo
  {volume} {53}},\ \bibinfo {pages} {722} (\bibinfo {year} {1984})}\BibitemShut
  {NoStop}%
\bibitem [{\citenamefont {Feldman}\ and\ \citenamefont
  {Halperin}(2021)}]{Feldman_2021}%
  \BibitemOpen
  \bibfield  {author} {\bibinfo {author} {\bibfnamefont {D.~E.}\ \bibnamefont
  {Feldman}}\ and\ \bibinfo {author} {\bibfnamefont {B.~I.}\ \bibnamefont
  {Halperin}},\ }\href {\doibase 10.1088/1361-6633/ac03aa} {\bibfield
  {journal} {\bibinfo  {journal} {Reports on Progress in Physics}\ }\textbf
  {\bibinfo {volume} {84}},\ \bibinfo {pages} {076501} (\bibinfo {year}
  {2021})}\BibitemShut {NoStop}%
\bibitem [{\citenamefont {Bartolomei}\ \emph {et~al.}(2020)\citenamefont
  {Bartolomei}, \citenamefont {Kumar}, \citenamefont {Bisognin}, \citenamefont
  {Marguerite}, \citenamefont {Berroir}, \citenamefont {Bocquillon},
  \citenamefont {Pla{\c{c}}ais}, \citenamefont {Cavanna}, \citenamefont {Dong},
  \citenamefont {Gennser} \emph {et~al.}}]{Bartolomei_2020}%
  \BibitemOpen
  \bibfield  {author} {\bibinfo {author} {\bibfnamefont {H.}~\bibnamefont
  {Bartolomei}}, \bibinfo {author} {\bibfnamefont {M.}~\bibnamefont {Kumar}},
  \bibinfo {author} {\bibfnamefont {R.}~\bibnamefont {Bisognin}}, \bibinfo
  {author} {\bibfnamefont {A.}~\bibnamefont {Marguerite}}, \bibinfo {author}
  {\bibfnamefont {J.-M.}\ \bibnamefont {Berroir}}, \bibinfo {author}
  {\bibfnamefont {E.}~\bibnamefont {Bocquillon}}, \bibinfo {author}
  {\bibfnamefont {B.}~\bibnamefont {Pla{\c{c}}ais}}, \bibinfo {author}
  {\bibfnamefont {A.}~\bibnamefont {Cavanna}}, \bibinfo {author} {\bibfnamefont
  {Q.}~\bibnamefont {Dong}}, \bibinfo {author} {\bibfnamefont {U.}~\bibnamefont
  {Gennser}},  \emph {et~al.},\ }\href@noop {} {\bibfield  {journal} {\bibinfo
  {journal} {Science}\ }\textbf {\bibinfo {volume} {368}},\ \bibinfo {pages}
  {173} (\bibinfo {year} {2020})}\BibitemShut {NoStop}%
\bibitem [{\citenamefont {Safi}\ \emph {et~al.}(2001)\citenamefont {Safi},
  \citenamefont {Devillard},\ and\ \citenamefont
  {Martin}}]{PhysRevLett.86.4628}%
  \BibitemOpen
  \bibfield  {author} {\bibinfo {author} {\bibfnamefont {I.}~\bibnamefont
  {Safi}}, \bibinfo {author} {\bibfnamefont {P.}~\bibnamefont {Devillard}}, \
  and\ \bibinfo {author} {\bibfnamefont {T.}~\bibnamefont {Martin}},\ }\href
  {\doibase 10.1103/PhysRevLett.86.4628} {\bibfield  {journal} {\bibinfo
  {journal} {Phys. Rev. Lett.}\ }\textbf {\bibinfo {volume} {86}},\ \bibinfo
  {pages} {4628} (\bibinfo {year} {2001})}\BibitemShut {NoStop}%
\bibitem [{\citenamefont {Campagnano}\ \emph {et~al.}(2012)\citenamefont
  {Campagnano}, \citenamefont {Zilberberg}, \citenamefont {Gornyi},
  \citenamefont {Feldman}, \citenamefont {Potter},\ and\ \citenamefont
  {Gefen}}]{PhysRevLett.109.106802}%
  \BibitemOpen
  \bibfield  {author} {\bibinfo {author} {\bibfnamefont {G.}~\bibnamefont
  {Campagnano}}, \bibinfo {author} {\bibfnamefont {O.}~\bibnamefont
  {Zilberberg}}, \bibinfo {author} {\bibfnamefont {I.~V.}\ \bibnamefont
  {Gornyi}}, \bibinfo {author} {\bibfnamefont {D.~E.}\ \bibnamefont {Feldman}},
  \bibinfo {author} {\bibfnamefont {A.~C.}\ \bibnamefont {Potter}}, \ and\
  \bibinfo {author} {\bibfnamefont {Y.}~\bibnamefont {Gefen}},\ }\href
  {\doibase 10.1103/PhysRevLett.109.106802} {\bibfield  {journal} {\bibinfo
  {journal} {Phys. Rev. Lett.}\ }\textbf {\bibinfo {volume} {109}},\ \bibinfo
  {pages} {106802} (\bibinfo {year} {2012})}\BibitemShut {NoStop}%
\bibitem [{\citenamefont {Campagnano}\ \emph {et~al.}(2013)\citenamefont
  {Campagnano}, \citenamefont {Zilberberg}, \citenamefont {Gornyi},\ and\
  \citenamefont {Gefen}}]{PhysRevB.88.235415}%
  \BibitemOpen
  \bibfield  {author} {\bibinfo {author} {\bibfnamefont {G.}~\bibnamefont
  {Campagnano}}, \bibinfo {author} {\bibfnamefont {O.}~\bibnamefont
  {Zilberberg}}, \bibinfo {author} {\bibfnamefont {I.~V.}\ \bibnamefont
  {Gornyi}}, \ and\ \bibinfo {author} {\bibfnamefont {Y.}~\bibnamefont
  {Gefen}},\ }\href {\doibase 10.1103/PhysRevB.88.235415} {\bibfield  {journal}
  {\bibinfo  {journal} {Phys. Rev. B}\ }\textbf {\bibinfo {volume} {88}},\
  \bibinfo {pages} {235415} (\bibinfo {year} {2013})}\BibitemShut {NoStop}%
\bibitem [{\citenamefont {Rosenow}\ \emph {et~al.}(2016)\citenamefont
  {Rosenow}, \citenamefont {Levkivskyi},\ and\ \citenamefont
  {Halperin}}]{PhysRevLett.116.156802}%
  \BibitemOpen
  \bibfield  {author} {\bibinfo {author} {\bibfnamefont {B.}~\bibnamefont
  {Rosenow}}, \bibinfo {author} {\bibfnamefont {I.~P.}\ \bibnamefont
  {Levkivskyi}}, \ and\ \bibinfo {author} {\bibfnamefont {B.~I.}\ \bibnamefont
  {Halperin}},\ }\href {\doibase 10.1103/PhysRevLett.116.156802} {\bibfield
  {journal} {\bibinfo  {journal} {Phys. Rev. Lett.}\ }\textbf {\bibinfo
  {volume} {116}},\ \bibinfo {pages} {156802} (\bibinfo {year}
  {2016})}\BibitemShut {NoStop}%
\bibitem [{\citenamefont {Nakamura}\ \emph {et~al.}(2020)\citenamefont
  {Nakamura}, \citenamefont {Liang}, \citenamefont {Gardner},\ and\
  \citenamefont {Manfra}}]{nakamura2020direct}%
  \BibitemOpen
  \bibfield  {author} {\bibinfo {author} {\bibfnamefont {J.}~\bibnamefont
  {Nakamura}}, \bibinfo {author} {\bibfnamefont {S.}~\bibnamefont {Liang}},
  \bibinfo {author} {\bibfnamefont {G.}~\bibnamefont {Gardner}}, \ and\
  \bibinfo {author} {\bibfnamefont {M.}~\bibnamefont {Manfra}},\ }\href
  {\doibase 10.1038/s41567-020-1019-1} {\bibfield  {journal} {\bibinfo
  {journal} {Nature Physics}\ }\textbf {\bibinfo {volume} {16}},\ \bibinfo
  {pages} {931} (\bibinfo {year} {2020})}\BibitemShut {NoStop}%
\bibitem [{\citenamefont {de~C.~Chamon}\ \emph {et~al.}(1997)\citenamefont
  {de~C.~Chamon}, \citenamefont {Freed}, \citenamefont {Kivelson},
  \citenamefont {Sondhi},\ and\ \citenamefont {Wen}}]{PhysRevB.55.2331}%
  \BibitemOpen
  \bibfield  {author} {\bibinfo {author} {\bibfnamefont {C.}~\bibnamefont
  {de~C.~Chamon}}, \bibinfo {author} {\bibfnamefont {D.~E.}\ \bibnamefont
  {Freed}}, \bibinfo {author} {\bibfnamefont {S.~A.}\ \bibnamefont {Kivelson}},
  \bibinfo {author} {\bibfnamefont {S.~L.}\ \bibnamefont {Sondhi}}, \ and\
  \bibinfo {author} {\bibfnamefont {X.~G.}\ \bibnamefont {Wen}},\ }\href
  {\doibase 10.1103/PhysRevB.55.2331} {\bibfield  {journal} {\bibinfo
  {journal} {Phys. Rev. B}\ }\textbf {\bibinfo {volume} {55}},\ \bibinfo
  {pages} {2331} (\bibinfo {year} {1997})}\BibitemShut {NoStop}%
\bibitem [{\citenamefont {Stern}\ and\ \citenamefont
  {Halperin}(2006)}]{PhysRevLett.96.016802}%
  \BibitemOpen
  \bibfield  {author} {\bibinfo {author} {\bibfnamefont {A.}~\bibnamefont
  {Stern}}\ and\ \bibinfo {author} {\bibfnamefont {B.~I.}\ \bibnamefont
  {Halperin}},\ }\href {\doibase 10.1103/PhysRevLett.96.016802} {\bibfield
  {journal} {\bibinfo  {journal} {Phys. Rev. Lett.}\ }\textbf {\bibinfo
  {volume} {96}},\ \bibinfo {pages} {016802} (\bibinfo {year}
  {2006})}\BibitemShut {NoStop}%
\bibitem [{\citenamefont {Bonderson}\ \emph {et~al.}(2006)\citenamefont
  {Bonderson}, \citenamefont {Kitaev},\ and\ \citenamefont
  {Shtengel}}]{PhysRevLett.96.016803}%
  \BibitemOpen
  \bibfield  {author} {\bibinfo {author} {\bibfnamefont {P.}~\bibnamefont
  {Bonderson}}, \bibinfo {author} {\bibfnamefont {A.}~\bibnamefont {Kitaev}}, \
  and\ \bibinfo {author} {\bibfnamefont {K.}~\bibnamefont {Shtengel}},\ }\href
  {\doibase 10.1103/PhysRevLett.96.016803} {\bibfield  {journal} {\bibinfo
  {journal} {Phys. Rev. Lett.}\ }\textbf {\bibinfo {volume} {96}},\ \bibinfo
  {pages} {016803} (\bibinfo {year} {2006})}\BibitemShut {NoStop}%
\bibitem [{\citenamefont {Grosfeld}\ \emph {et~al.}(2006)\citenamefont
  {Grosfeld}, \citenamefont {Simon},\ and\ \citenamefont
  {Stern}}]{PhysRevLett.96.226803}%
  \BibitemOpen
  \bibfield  {author} {\bibinfo {author} {\bibfnamefont {E.}~\bibnamefont
  {Grosfeld}}, \bibinfo {author} {\bibfnamefont {S.~H.}\ \bibnamefont {Simon}},
  \ and\ \bibinfo {author} {\bibfnamefont {A.}~\bibnamefont {Stern}},\ }\href
  {\doibase 10.1103/PhysRevLett.96.226803} {\bibfield  {journal} {\bibinfo
  {journal} {Phys. Rev. Lett.}\ }\textbf {\bibinfo {volume} {96}},\ \bibinfo
  {pages} {226803} (\bibinfo {year} {2006})}\BibitemShut {NoStop}%
\bibitem [{\citenamefont {Law}\ \emph {et~al.}(2006)\citenamefont {Law},
  \citenamefont {Feldman},\ and\ \citenamefont {Gefen}}]{PhysRevB.74.045319}%
  \BibitemOpen
  \bibfield  {author} {\bibinfo {author} {\bibfnamefont {K.~T.}\ \bibnamefont
  {Law}}, \bibinfo {author} {\bibfnamefont {D.~E.}\ \bibnamefont {Feldman}}, \
  and\ \bibinfo {author} {\bibfnamefont {Y.}~\bibnamefont {Gefen}},\ }\href
  {\doibase 10.1103/PhysRevB.74.045319} {\bibfield  {journal} {\bibinfo
  {journal} {Phys. Rev. B}\ }\textbf {\bibinfo {volume} {74}},\ \bibinfo
  {pages} {045319} (\bibinfo {year} {2006})}\BibitemShut {NoStop}%
\bibitem [{\citenamefont {Kim}(2006)}]{PhysRevLett.97.216404}%
  \BibitemOpen
  \bibfield  {author} {\bibinfo {author} {\bibfnamefont {E.-A.}\ \bibnamefont
  {Kim}},\ }\href {\doibase 10.1103/PhysRevLett.97.216404} {\bibfield
  {journal} {\bibinfo  {journal} {Phys. Rev. Lett.}\ }\textbf {\bibinfo
  {volume} {97}},\ \bibinfo {pages} {216404} (\bibinfo {year}
  {2006})}\BibitemShut {NoStop}%
\bibitem [{\citenamefont {Feldman}\ \emph {et~al.}(2007)\citenamefont
  {Feldman}, \citenamefont {Gefen}, \citenamefont {Kitaev}, \citenamefont
  {Law},\ and\ \citenamefont {Stern}}]{PhysRevB.76.085333}%
  \BibitemOpen
  \bibfield  {author} {\bibinfo {author} {\bibfnamefont {D.~E.}\ \bibnamefont
  {Feldman}}, \bibinfo {author} {\bibfnamefont {Y.}~\bibnamefont {Gefen}},
  \bibinfo {author} {\bibfnamefont {A.}~\bibnamefont {Kitaev}}, \bibinfo
  {author} {\bibfnamefont {K.~T.}\ \bibnamefont {Law}}, \ and\ \bibinfo
  {author} {\bibfnamefont {A.}~\bibnamefont {Stern}},\ }\href {\doibase
  10.1103/PhysRevB.76.085333} {\bibfield  {journal} {\bibinfo  {journal} {Phys.
  Rev. B}\ }\textbf {\bibinfo {volume} {76}},\ \bibinfo {pages} {085333}
  (\bibinfo {year} {2007})}\BibitemShut {NoStop}%
\bibitem [{\citenamefont {Rosenow}\ and\ \citenamefont
  {Stern}(2020)}]{PhysRevLett.124.106805}%
  \BibitemOpen
  \bibfield  {author} {\bibinfo {author} {\bibfnamefont {B.}~\bibnamefont
  {Rosenow}}\ and\ \bibinfo {author} {\bibfnamefont {A.}~\bibnamefont
  {Stern}},\ }\href {\doibase 10.1103/PhysRevLett.124.106805} {\bibfield
  {journal} {\bibinfo  {journal} {Phys. Rev. Lett.}\ }\textbf {\bibinfo
  {volume} {124}},\ \bibinfo {pages} {106805} (\bibinfo {year}
  {2020})}\BibitemShut {NoStop}%
\bibitem [{\citenamefont {{Venkatachalam}}\ \emph {et~al.}(2012)\citenamefont
  {{Venkatachalam}}, \citenamefont {{Hart}}, \citenamefont {{Pfeiffer}},
  \citenamefont {{West}},\ and\ \citenamefont
  {{Yacoby}}}]{2012NatPh...8..676V}%
  \BibitemOpen
  \bibfield  {author} {\bibinfo {author} {\bibfnamefont {V.}~\bibnamefont
  {{Venkatachalam}}}, \bibinfo {author} {\bibfnamefont {S.}~\bibnamefont
  {{Hart}}}, \bibinfo {author} {\bibfnamefont {L.}~\bibnamefont {{Pfeiffer}}},
  \bibinfo {author} {\bibfnamefont {K.}~\bibnamefont {{West}}}, \ and\ \bibinfo
  {author} {\bibfnamefont {A.}~\bibnamefont {{Yacoby}}},\ }\href {\doibase
  10.1038/nphys2384} {\bibfield  {journal} {\bibinfo  {journal} {Nature
  Physics}\ }\textbf {\bibinfo {volume} {8}},\ \bibinfo {pages} {676} (\bibinfo
  {year} {2012})}\BibitemShut {NoStop}%
\bibitem [{\citenamefont {{Bid}}\ \emph {et~al.}(2010)\citenamefont {{Bid}},
  \citenamefont {{Ofek}}, \citenamefont {{Inoue}}, \citenamefont {{Heiblum}},
  \citenamefont {{Kane}}, \citenamefont {{Umansky}},\ and\ \citenamefont
  {{Mahalu}}}]{2010Natur.466..585B}%
  \BibitemOpen
  \bibfield  {author} {\bibinfo {author} {\bibfnamefont {A.}~\bibnamefont
  {{Bid}}}, \bibinfo {author} {\bibfnamefont {N.}~\bibnamefont {{Ofek}}},
  \bibinfo {author} {\bibfnamefont {H.}~\bibnamefont {{Inoue}}}, \bibinfo
  {author} {\bibfnamefont {M.}~\bibnamefont {{Heiblum}}}, \bibinfo {author}
  {\bibfnamefont {C.~L.}\ \bibnamefont {{Kane}}}, \bibinfo {author}
  {\bibfnamefont {V.}~\bibnamefont {{Umansky}}}, \ and\ \bibinfo {author}
  {\bibfnamefont {D.}~\bibnamefont {{Mahalu}}},\ }\href {\doibase
  10.1038/nature09277} {\bibfield  {journal} {\bibinfo  {journal} {Nature}\
  }\textbf {\bibinfo {volume} {466}},\ \bibinfo {pages} {585} (\bibinfo {year}
  {2010})}\BibitemShut {NoStop}%
\bibitem [{\citenamefont {Dolev}\ \emph {et~al.}(2011)\citenamefont {Dolev},
  \citenamefont {Gross}, \citenamefont {Sabo}, \citenamefont {Gurman},
  \citenamefont {Heiblum}, \citenamefont {Umansky},\ and\ \citenamefont
  {Mahalu}}]{PhysRevLett.107.036805}%
  \BibitemOpen
  \bibfield  {author} {\bibinfo {author} {\bibfnamefont {M.}~\bibnamefont
  {Dolev}}, \bibinfo {author} {\bibfnamefont {Y.}~\bibnamefont {Gross}},
  \bibinfo {author} {\bibfnamefont {R.}~\bibnamefont {Sabo}}, \bibinfo {author}
  {\bibfnamefont {I.}~\bibnamefont {Gurman}}, \bibinfo {author} {\bibfnamefont
  {M.}~\bibnamefont {Heiblum}}, \bibinfo {author} {\bibfnamefont
  {V.}~\bibnamefont {Umansky}}, \ and\ \bibinfo {author} {\bibfnamefont
  {D.}~\bibnamefont {Mahalu}},\ }\href {\doibase
  10.1103/PhysRevLett.107.036805} {\bibfield  {journal} {\bibinfo  {journal}
  {Phys. Rev. Lett.}\ }\textbf {\bibinfo {volume} {107}},\ \bibinfo {pages}
  {036805} (\bibinfo {year} {2011})}\BibitemShut {NoStop}%
\bibitem [{\citenamefont {Gross}\ \emph {et~al.}(2012)\citenamefont {Gross},
  \citenamefont {Dolev}, \citenamefont {Heiblum}, \citenamefont {Umansky},\
  and\ \citenamefont {Mahalu}}]{PhysRevLett.108.226801}%
  \BibitemOpen
  \bibfield  {author} {\bibinfo {author} {\bibfnamefont {Y.}~\bibnamefont
  {Gross}}, \bibinfo {author} {\bibfnamefont {M.}~\bibnamefont {Dolev}},
  \bibinfo {author} {\bibfnamefont {M.}~\bibnamefont {Heiblum}}, \bibinfo
  {author} {\bibfnamefont {V.}~\bibnamefont {Umansky}}, \ and\ \bibinfo
  {author} {\bibfnamefont {D.}~\bibnamefont {Mahalu}},\ }\href {\doibase
  10.1103/PhysRevLett.108.226801} {\bibfield  {journal} {\bibinfo  {journal}
  {Phys. Rev. Lett.}\ }\textbf {\bibinfo {volume} {108}},\ \bibinfo {pages}
  {226801} (\bibinfo {year} {2012})}\BibitemShut {NoStop}%
\bibitem [{\citenamefont {{Gurman}}\ \emph {et~al.}(2012)\citenamefont
  {{Gurman}}, \citenamefont {{Sabo}}, \citenamefont {{Heiblum}}, \citenamefont
  {{Umansky}},\ and\ \citenamefont {{Mahalu}}}]{2012NatCo...3E1289G}%
  \BibitemOpen
  \bibfield  {author} {\bibinfo {author} {\bibfnamefont {I.}~\bibnamefont
  {{Gurman}}}, \bibinfo {author} {\bibfnamefont {R.}~\bibnamefont {{Sabo}}},
  \bibinfo {author} {\bibfnamefont {M.}~\bibnamefont {{Heiblum}}}, \bibinfo
  {author} {\bibfnamefont {V.}~\bibnamefont {{Umansky}}}, \ and\ \bibinfo
  {author} {\bibfnamefont {D.}~\bibnamefont {{Mahalu}}},\ }\href {\doibase
  10.1038/ncomms2305} {\bibfield  {journal} {\bibinfo  {journal} {Nature
  Communications}\ }\textbf {\bibinfo {volume} {3}},\ \bibinfo {eid} {1289}
  (\bibinfo {year} {2012})}\BibitemShut {NoStop}%
\bibitem [{\citenamefont {{Inoue}}\ \emph {et~al.}(2014)\citenamefont
  {{Inoue}}, \citenamefont {{Grivnin}}, \citenamefont {{Ronen}}, \citenamefont
  {{Heiblum}}, \citenamefont {{Umansky}},\ and\ \citenamefont
  {{Mahalu}}}]{2014NatCo...5E4067I}%
  \BibitemOpen
  \bibfield  {author} {\bibinfo {author} {\bibfnamefont {H.}~\bibnamefont
  {{Inoue}}}, \bibinfo {author} {\bibfnamefont {A.}~\bibnamefont {{Grivnin}}},
  \bibinfo {author} {\bibfnamefont {Y.}~\bibnamefont {{Ronen}}}, \bibinfo
  {author} {\bibfnamefont {M.}~\bibnamefont {{Heiblum}}}, \bibinfo {author}
  {\bibfnamefont {V.}~\bibnamefont {{Umansky}}}, \ and\ \bibinfo {author}
  {\bibfnamefont {D.}~\bibnamefont {{Mahalu}}},\ }\href {\doibase
  10.1038/ncomms5067} {\bibfield  {journal} {\bibinfo  {journal} {Nature
  Communications}\ }\textbf {\bibinfo {volume} {5}},\ \bibinfo {eid} {4067}
  (\bibinfo {year} {2014})}\BibitemShut {NoStop}%
\bibitem [{\citenamefont {Kane}\ \emph {et~al.}(1994)\citenamefont {Kane},
  \citenamefont {Fisher},\ and\ \citenamefont
  {Polchinski}}]{kane_randomness_1994}%
  \BibitemOpen
  \bibfield  {author} {\bibinfo {author} {\bibfnamefont {C.~L.}\ \bibnamefont
  {Kane}}, \bibinfo {author} {\bibfnamefont {M.~P.~A.}\ \bibnamefont {Fisher}},
  \ and\ \bibinfo {author} {\bibfnamefont {J.}~\bibnamefont {Polchinski}},\
  }\href {http://link.aps.org/doi/10.1103/PhysRevLett.72.4129} {\bibfield
  {journal} {\bibinfo  {journal} {Phys. Rev. Lett.}\ }\textbf {\bibinfo
  {volume} {72}},\ \bibinfo {pages} {4129} (\bibinfo {year}
  {1994})}\BibitemShut {NoStop}%
\bibitem [{\citenamefont {{MacDonald}}(1990)}]{1990PhRvL..64..220M}%
  \BibitemOpen
  \bibfield  {author} {\bibinfo {author} {\bibfnamefont {A.~H.}\ \bibnamefont
  {{MacDonald}}},\ }\href {\doibase 10.1103/PhysRevLett.64.220} {\bibfield
  {journal} {\bibinfo  {journal} {Phys. Rev. Lett.}\ }\textbf {\bibinfo
  {volume} {64}},\ \bibinfo {pages} {220} (\bibinfo {year} {1990})}\BibitemShut
  {NoStop}%
\bibitem [{\citenamefont {Wen}(1990)}]{PhysRevLett.64.2206}%
  \BibitemOpen
  \bibfield  {author} {\bibinfo {author} {\bibfnamefont {X.~G.}\ \bibnamefont
  {Wen}},\ }\href {\doibase 10.1103/PhysRevLett.64.2206} {\bibfield  {journal}
  {\bibinfo  {journal} {Phys. Rev. Lett.}\ }\textbf {\bibinfo {volume} {64}},\
  \bibinfo {pages} {2206} (\bibinfo {year} {1990})}\BibitemShut {NoStop}%
\bibitem [{\citenamefont {Wang}\ \emph {et~al.}(2013)\citenamefont {Wang},
  \citenamefont {Meir},\ and\ \citenamefont {Gefen}}]{wang_edge_2013}%
  \BibitemOpen
  \bibfield  {author} {\bibinfo {author} {\bibfnamefont {J.}~\bibnamefont
  {Wang}}, \bibinfo {author} {\bibfnamefont {Y.}~\bibnamefont {Meir}}, \ and\
  \bibinfo {author} {\bibfnamefont {Y.}~\bibnamefont {Gefen}},\ }\href
  {http://link.aps.org/doi/10.1103/PhysRevLett.111.246803} {\bibfield
  {journal} {\bibinfo  {journal} {Phys. Rev. Lett.}\ }\textbf {\bibinfo
  {volume} {111}},\ \bibinfo {pages} {246803} (\bibinfo {year}
  {2013})}\BibitemShut {NoStop}%
\bibitem [{\citenamefont {Wen}(2004)}]{wen_quantum_2004}%
  \BibitemOpen
  \bibfield  {author} {\bibinfo {author} {\bibfnamefont {X.-G.}\ \bibnamefont
  {Wen}},\ }\href@noop {} {\emph {\bibinfo {title} {Quantum {Field} {Theory} of
  {Many}-{Body} {Systems}}}}\ (\bibinfo  {publisher} {Oxford University
  Press},\ \bibinfo {year} {2004})\BibitemShut {NoStop}%
\bibitem [{\citenamefont {{Ronen}}\ \emph {et~al.}(2018)\citenamefont
  {{Ronen}}, \citenamefont {{Cohen}}, \citenamefont {{Banitt}}, \citenamefont
  {{Heiblum}},\ and\ \citenamefont {{Umansky}}}]{2018NatPh..14..411R}%
  \BibitemOpen
  \bibfield  {author} {\bibinfo {author} {\bibfnamefont {Y.}~\bibnamefont
  {{Ronen}}}, \bibinfo {author} {\bibfnamefont {Y.}~\bibnamefont {{Cohen}}},
  \bibinfo {author} {\bibfnamefont {D.}~\bibnamefont {{Banitt}}}, \bibinfo
  {author} {\bibfnamefont {M.}~\bibnamefont {{Heiblum}}}, \ and\ \bibinfo
  {author} {\bibfnamefont {V.}~\bibnamefont {{Umansky}}},\ }\href {\doibase
  10.1038/s41567-017-0035-2} {\bibfield  {journal} {\bibinfo  {journal} {Nat.
  Phys.}\ }\textbf {\bibinfo {volume} {14}},\ \bibinfo {pages} {411} (\bibinfo
  {year} {2018})}\BibitemShut {NoStop}%
\bibitem [{\citenamefont {Wang}\ \emph {et~al.}(2021)\citenamefont {Wang},
  \citenamefont {Ponomarenko}, \citenamefont {Wan}, \citenamefont {West},
  \citenamefont {Baldwin}, \citenamefont {Pfeiffer}, \citenamefont
  {Lyanda-Geller},\ and\ \citenamefont {Rokhinson}}]{wang2021transport}%
  \BibitemOpen
  \bibfield  {author} {\bibinfo {author} {\bibfnamefont {Y.}~\bibnamefont
  {Wang}}, \bibinfo {author} {\bibfnamefont {V.}~\bibnamefont {Ponomarenko}},
  \bibinfo {author} {\bibfnamefont {Z.}~\bibnamefont {Wan}}, \bibinfo {author}
  {\bibfnamefont {K.~W.}\ \bibnamefont {West}}, \bibinfo {author}
  {\bibfnamefont {K.~W.}\ \bibnamefont {Baldwin}}, \bibinfo {author}
  {\bibfnamefont {L.~N.}\ \bibnamefont {Pfeiffer}}, \bibinfo {author}
  {\bibfnamefont {Y.}~\bibnamefont {Lyanda-Geller}}, \ and\ \bibinfo {author}
  {\bibfnamefont {L.~P.}\ \bibnamefont {Rokhinson}},\ }\href
  {http://dx.doi.org/10.1038/s41467-021-25631-2} {\bibfield  {journal}
  {\bibinfo  {journal} {Nature Communications}\ }\textbf {\bibinfo {volume}
  {12}} (\bibinfo {year} {2021})}\BibitemShut {NoStop}%
\bibitem [{Note1()}]{Note1}%
  \BibitemOpen
  \bibinfo {note} {We note that neutralons are semions, hence the allowed
  backscattering process involves four neutralons, making it highly irrelevant
  at low energies~\protect \citep {kane_randomness_1994,Kane_1997}. (The
  scaling dimension of the neutralon backscattering operator is
  2.)}\BibitemShut {NoStop}%
\bibitem [{\citenamefont {Kamenev}\ and\ \citenamefont
  {Gefen}(2015)}]{PhysRevLett.114.156401}%
  \BibitemOpen
  \bibfield  {author} {\bibinfo {author} {\bibfnamefont {A.}~\bibnamefont
  {Kamenev}}\ and\ \bibinfo {author} {\bibfnamefont {Y.}~\bibnamefont
  {Gefen}},\ }\href {\doibase 10.1103/PhysRevLett.114.156401} {\bibfield
  {journal} {\bibinfo  {journal} {Phys. Rev. Lett.}\ }\textbf {\bibinfo
  {volume} {114}},\ \bibinfo {pages} {156401} (\bibinfo {year}
  {2015})}\BibitemShut {NoStop}%
\bibitem [{\citenamefont {Bhattacharyya}\ \emph {et~al.}(2019)\citenamefont
  {Bhattacharyya}, \citenamefont {Banerjee}, \citenamefont {Heiblum},
  \citenamefont {Mahalu},\ and\ \citenamefont
  {Umansky}}]{PhysRevLett.122.246801}%
  \BibitemOpen
  \bibfield  {author} {\bibinfo {author} {\bibfnamefont {R.}~\bibnamefont
  {Bhattacharyya}}, \bibinfo {author} {\bibfnamefont {M.}~\bibnamefont
  {Banerjee}}, \bibinfo {author} {\bibfnamefont {M.}~\bibnamefont {Heiblum}},
  \bibinfo {author} {\bibfnamefont {D.}~\bibnamefont {Mahalu}}, \ and\ \bibinfo
  {author} {\bibfnamefont {V.}~\bibnamefont {Umansky}},\ }\href {\doibase
  10.1103/PhysRevLett.122.246801} {\bibfield  {journal} {\bibinfo  {journal}
  {Phys. Rev. Lett.}\ }\textbf {\bibinfo {volume} {122}},\ \bibinfo {pages}
  {246801} (\bibinfo {year} {2019})}\BibitemShut {NoStop}%
\bibitem [{\citenamefont {Goldstein}\ and\ \citenamefont
  {Gefen}(2016)}]{PhysRevLett.117.276804}%
  \BibitemOpen
  \bibfield  {author} {\bibinfo {author} {\bibfnamefont {M.}~\bibnamefont
  {Goldstein}}\ and\ \bibinfo {author} {\bibfnamefont {Y.}~\bibnamefont
  {Gefen}},\ }\href {\doibase 10.1103/PhysRevLett.117.276804} {\bibfield
  {journal} {\bibinfo  {journal} {Phys. Rev. Lett.}\ }\textbf {\bibinfo
  {volume} {117}},\ \bibinfo {pages} {276804} (\bibinfo {year}
  {2016})}\BibitemShut {NoStop}%
\bibitem [{Note2()}]{Note2}%
  \BibitemOpen
  \bibinfo {note} {See Supplemental Material, which includes Refs.~\protect
  \citep {PhysRevB.84.195436,PhysRevB.36.4581}.}\BibitemShut {Stop}%
\bibitem [{Note3()}]{Note3}%
  \BibitemOpen
  \bibinfo {note} {The chiral fields obey $[\phi _{i}(x),\phi _{j}(x')]=\pi
  i(\protect \mathbf {K}^{-1})_{ij}\protect \text {sgn}(x-x')$~\protect \citep
  {2017AnPhy.385..287P}.}\BibitemShut {Stop}%
\bibitem [{\citenamefont {Giamarchi}\ and\ \citenamefont
  {Schulz}(1988)}]{giamarchi_anderson_1988}%
  \BibitemOpen
  \bibfield  {author} {\bibinfo {author} {\bibfnamefont {T.}~\bibnamefont
  {Giamarchi}}\ and\ \bibinfo {author} {\bibfnamefont {H.~J.}\ \bibnamefont
  {Schulz}},\ }\href {http://link.aps.org/doi/10.1103/PhysRevB.37.325}
  {\bibfield  {journal} {\bibinfo  {journal} {Phys. Rev. B}\ }\textbf {\bibinfo
  {volume} {37}},\ \bibinfo {pages} {325} (\bibinfo {year} {1988})}\BibitemShut
  {NoStop}%
\bibitem [{Note4()}]{Note4}%
  \BibitemOpen
  \bibinfo {note} {We may estimate $D_{\protect \text {KFP}}$ by using the RG
  equation $dW/d\protect \qopname \relax o{ln}D^{-1}=(3-2\delta )W$ where
  $\delta <3/2$ is the scaling dimension and $D$ the reduced bandwidth. At
  strong coupling we have $W(D_{\protect \text {KFP}})\sim v_{0}^{2}$ which
  yields $D_{\protect \text {KFP}}\sim D_{0}(W(D_{0})/D_{0})^{1/(3-2\delta )}$
  in terms of the bare bandwidth $D_{0}$. If the KFP fixed point is not fully
  reached (say, at temperature $T\gtrsim D_{\protect \text {KFP}}$) there will
  be RG irrelevant interactions such as $\partial _{x}\phi _{+,2/3,\downarrow
  }\partial _{x}\phi _{-,0}$ between the charge and neutral modes.}\BibitemShut
  {Stop}%
\bibitem [{\citenamefont {Giamarchi}(2003)}]{giamarchi_quantum_2003}%
  \BibitemOpen
  \bibfield  {author} {\bibinfo {author} {\bibfnamefont {T.}~\bibnamefont
  {Giamarchi}},\ }\href@noop {} {\emph {\bibinfo {title} {Quantum {Physics} in
  {One} {Dimension}}}},\ International {Series} of {Monographs} on {Physics}\
  (\bibinfo  {publisher} {Clarendon Press},\ \bibinfo {year}
  {2003})\BibitemShut {NoStop}%
\bibitem [{Note5()}]{Note5}%
  \BibitemOpen
  \bibinfo {note} {For a range of bare parameters, attraction may develop
  between the charge modes, in which case pairing becomes relevant~\protect
  \citep {Vayrynen_2019}. Here we assume that this is not the
  case.}\BibitemShut {Stop}%
\bibitem [{\citenamefont {Levin}(2013)}]{PhysRevX.3.021009}%
  \BibitemOpen
  \bibfield  {author} {\bibinfo {author} {\bibfnamefont {M.}~\bibnamefont
  {Levin}},\ }\href {\doibase 10.1103/PhysRevX.3.021009} {\bibfield  {journal}
  {\bibinfo  {journal} {Phys. Rev. X}\ }\textbf {\bibinfo {volume} {3}},\
  \bibinfo {pages} {021009} (\bibinfo {year} {2013})}\BibitemShut {NoStop}%
\bibitem [{\citenamefont {{Protopopov}}\ \emph {et~al.}(2017)\citenamefont
  {{Protopopov}}, \citenamefont {{Gefen}},\ and\ \citenamefont
  {{Mirlin}}}]{2017AnPhy.385..287P}%
  \BibitemOpen
  \bibfield  {author} {\bibinfo {author} {\bibfnamefont {I.~V.}\ \bibnamefont
  {{Protopopov}}}, \bibinfo {author} {\bibfnamefont {Y.}~\bibnamefont
  {{Gefen}}}, \ and\ \bibinfo {author} {\bibfnamefont {A.~D.}\ \bibnamefont
  {{Mirlin}}},\ }\href {\doibase 10.1016/j.aop.2017.07.015} {\bibfield
  {journal} {\bibinfo  {journal} {Annals of Physics}\ }\textbf {\bibinfo
  {volume} {385}},\ \bibinfo {pages} {287} (\bibinfo {year}
  {2017})}\BibitemShut {NoStop}%
\bibitem [{\citenamefont {Gogolin}\ \emph {et~al.}(2004)\citenamefont
  {Gogolin}, \citenamefont {Nersesyan},\ and\ \citenamefont
  {Tsvelik}}]{gogolin2004bosonization}%
  \BibitemOpen
  \bibfield  {author} {\bibinfo {author} {\bibfnamefont {A.}~\bibnamefont
  {Gogolin}}, \bibinfo {author} {\bibfnamefont {A.}~\bibnamefont {Nersesyan}},
  \ and\ \bibinfo {author} {\bibfnamefont {A.}~\bibnamefont {Tsvelik}},\ }\href
  {https://books.google.com/books?id=j2AcohdUuaoC} {\emph {\bibinfo {title}
  {Bosonization and Strongly Correlated Systems}}}\ (\bibinfo  {publisher}
  {Cambridge University Press},\ \bibinfo {year} {2004})\BibitemShut {NoStop}%
\bibitem [{\citenamefont {Kane}\ and\ \citenamefont
  {Fisher}(1994)}]{PhysRevLett.72.724}%
  \BibitemOpen
  \bibfield  {author} {\bibinfo {author} {\bibfnamefont {C.~L.}\ \bibnamefont
  {Kane}}\ and\ \bibinfo {author} {\bibfnamefont {M.~P.~A.}\ \bibnamefont
  {Fisher}},\ }\href {\doibase 10.1103/PhysRevLett.72.724} {\bibfield
  {journal} {\bibinfo  {journal} {Phys. Rev. Lett.}\ }\textbf {\bibinfo
  {volume} {72}},\ \bibinfo {pages} {724} (\bibinfo {year} {1994})}\BibitemShut
  {NoStop}%
\bibitem [{\citenamefont {Bid}\ \emph {et~al.}(2009)\citenamefont {Bid},
  \citenamefont {Ofek}, \citenamefont {Heiblum}, \citenamefont {Umansky},\ and\
  \citenamefont {Mahalu}}]{PhysRevLett.103.236802}%
  \BibitemOpen
  \bibfield  {author} {\bibinfo {author} {\bibfnamefont {A.}~\bibnamefont
  {Bid}}, \bibinfo {author} {\bibfnamefont {N.}~\bibnamefont {Ofek}}, \bibinfo
  {author} {\bibfnamefont {M.}~\bibnamefont {Heiblum}}, \bibinfo {author}
  {\bibfnamefont {V.}~\bibnamefont {Umansky}}, \ and\ \bibinfo {author}
  {\bibfnamefont {D.}~\bibnamefont {Mahalu}},\ }\href {\doibase
  10.1103/PhysRevLett.103.236802} {\bibfield  {journal} {\bibinfo  {journal}
  {Phys. Rev. Lett.}\ }\textbf {\bibinfo {volume} {103}},\ \bibinfo {pages}
  {236802} (\bibinfo {year} {2009})}\BibitemShut {NoStop}%
\bibitem [{\citenamefont {Kane}\ and\ \citenamefont
  {Fisher}(1997)}]{Kane_1997}%
  \BibitemOpen
  \bibfield  {author} {\bibinfo {author} {\bibfnamefont {C.~L.}\ \bibnamefont
  {Kane}}\ and\ \bibinfo {author} {\bibfnamefont {M.~P.~A.}\ \bibnamefont
  {Fisher}},\ }\href {\doibase 10.1103/physrevb.56.15231} {\bibfield  {journal}
  {\bibinfo  {journal} {Phys. Rev. B}\ }\textbf {\bibinfo {volume} {56}},\
  \bibinfo {pages} {15231} (\bibinfo {year} {1997})}\BibitemShut {NoStop}%
\bibitem [{\citenamefont {Fidkowski}\ \emph {et~al.}(2011)\citenamefont
  {Fidkowski}, \citenamefont {Lutchyn}, \citenamefont {Nayak},\ and\
  \citenamefont {Fisher}}]{PhysRevB.84.195436}%
  \BibitemOpen
  \bibfield  {author} {\bibinfo {author} {\bibfnamefont {L.}~\bibnamefont
  {Fidkowski}}, \bibinfo {author} {\bibfnamefont {R.~M.}\ \bibnamefont
  {Lutchyn}}, \bibinfo {author} {\bibfnamefont {C.}~\bibnamefont {Nayak}}, \
  and\ \bibinfo {author} {\bibfnamefont {M.~P.~A.}\ \bibnamefont {Fisher}},\
  }\href {\doibase 10.1103/PhysRevB.84.195436} {\bibfield  {journal} {\bibinfo
  {journal} {Phys. Rev. B}\ }\textbf {\bibinfo {volume} {84}},\ \bibinfo
  {pages} {195436} (\bibinfo {year} {2011})}\BibitemShut {NoStop}%
\bibitem [{\citenamefont {Thouless}\ and\ \citenamefont
  {Li}(1987)}]{PhysRevB.36.4581}%
  \BibitemOpen
  \bibfield  {author} {\bibinfo {author} {\bibfnamefont {D.~J.}\ \bibnamefont
  {Thouless}}\ and\ \bibinfo {author} {\bibfnamefont {Q.}~\bibnamefont {Li}},\
  }\href {\doibase 10.1103/PhysRevB.36.4581} {\bibfield  {journal} {\bibinfo
  {journal} {Phys. Rev. B}\ }\textbf {\bibinfo {volume} {36}},\ \bibinfo
  {pages} {4581} (\bibinfo {year} {1987})}\BibitemShut {NoStop}%
\bibitem [{\citenamefont {V\"ayrynen}\ \emph {et~al.}(2019)\citenamefont
  {V\"ayrynen}, \citenamefont {Goldstein},\ and\ \citenamefont
  {Gefen}}]{Vayrynen_2019}%
  \BibitemOpen
  \bibfield  {author} {\bibinfo {author} {\bibfnamefont {J.~I.}\ \bibnamefont
  {V\"ayrynen}}, \bibinfo {author} {\bibfnamefont {M.}~\bibnamefont
  {Goldstein}}, \ and\ \bibinfo {author} {\bibfnamefont {Y.}~\bibnamefont
  {Gefen}},\ }\href {\doibase 10.1103/PhysRevLett.122.236802} {\bibfield
  {journal} {\bibinfo  {journal} {Phys. Rev. Lett.}\ }\textbf {\bibinfo
  {volume} {122}},\ \bibinfo {pages} {236802} (\bibinfo {year}
  {2019})}\BibitemShut {NoStop}%
\end{thebibliography}%


\newpage
~
\newpage

\setcounter{page}{1}

\renewcommand{\thesection}{SM\arabic{section}}
\renewcommand{\theequation}{S\arabic{equation}}
\renewcommand{\thefigure}{S\arabic{figure}}
\renewcommand{\bibnumfmt}[1]{[S#1]}

\begin{widetext}

\section{Supplementary Material to ``Superconductivity of neutral modes in quantum Hall edges''}

In this Supplementary Material, we present details of the actions~(\ref{eq:ShomogDiab})
and~(\ref{eq:SKFPFP}), the RG equations (\ref{eq:Rg1})--(\ref{Rg2}), the orthogonal transformation~(\ref{eq:JTransformation}), discuss the neutralon pairing term and its signatures in a quantum antidot, 
and show alternative QPC designs to Fig.~\ref{fig:QPC}. 

\subsection{V-matrices}

In this Section, we give explicit expressions for the $V$-matrices
introduced in the main text. We also discuss the lowest-order RG equations
for the pairing term, valid in the limit of relatively strong neutralon-neutralon
interaction. 

The $V$-matrix in Eq.~(\ref{eq:ShomogDiab}) is given by~\citep{kane_randomness_1994}
\begin{equation}
\mathbf{V}=\left(\begin{array}{cc}
\mathbf{v} & \mathbf{u}\\
\mathbf{u}^{T} & \mathbf{v}
\end{array}\right)\,,\quad\mathbf{v}=\left(\begin{array}{cc}
3v_{1/3} & v_{1/3,1}\\
v_{1/3,1} & v_{1}
\end{array}\right)\,,\quad\mathbf{u}=\left(\begin{array}{cc}
u_{1/3} & u_{1/3,1}\\
u_{1/3,1} & u_{1}
\end{array}\right)\,,
\end{equation}
where $v_{1/3}$ and $v_{1}$ are the velocities of the $\nu=1/3$
and $\nu=1$ modes and $v_{1/3,1}$ is their interaction strength.
We assume that these quantities are the same for both top and bottom
layers. The $2\times2$ matrix $\mathbf{u}$ characterizes the inter-layer
repulsive interactions; we assume its matrix elements are small in
comparison to the intralayer terms $\mathbf{v}$, and the RG flow
is analogous to the one in KFP theory~\citep{kane_randomness_1994}. 

At the KFP fixed point, in Eq.~(\ref{eq:SKFPFP}) we have 
\begin{equation}
\mathbf{V}_{\text{KFP}}=\left(\begin{array}{cc}
\left(\begin{array}{cc}
v_{2/3} & v_{2/3,2/3}\\
v_{2/3,2/3} & v_{2/3}
\end{array}\right) & 0\\
0 & \left(\begin{array}{cc}
v_{0} & v_{0,0}\\
v_{0,0} & v_{0}
\end{array}\right)
\end{array}\right)\,,
\end{equation}
where $v_{0}$ is the neutral mode velocity. Due to the randomness
in the neutral sector, the interactions that couple to the neutralons
are generally irrelevant perturbations and can be left out. We will
keep the neutralon-neutralon interaction $v_{0,0}$ which, in the
case of layer-correlated disorder, is not irrelevant. The bare value
of this interaction is $v_{0,0}=\frac{1}{2}(u_{1/3}+u_{1}-2u_{1/3,1})$
and can be positive (repulsion) or negative (attraction). In most
designs we have $1/3$ and $1$ modes closest together so that $u_{1/3,1}>u_{1},\,u_{1/3}$
and therefore $v_{0,0}<0$ would be expected. The neutralons are charge dipoles
so their attraction is not entirely surprising. The sign determines
the relevant gap opening perturbation in the neutral sector and $v_{0,0}<0$
makes pairing the relevant perturbation. In the charge sector, $v_{2/3}$
is the velocity of both $\nu=2/3$ charge modes and $v_{2/3,2/3}$
denotes their interaction strength. {[}Like $v_{0,0}$, also $v_{2/3,2/3}$
can be obtained from the matrix $\mathbf{u}$ with the help of Eq.~(\ref{eq:neutralcharge}) of the main text.{]}

We can diagonalize the neutral sector $V$-matrix with the rotation
\begin{equation}
\left(\!\begin{array}{c}
\phi_{+,0}\\
\phi_{-,0}
\end{array}\!\right)\negmedspace=\negmedspace\left(\!\begin{array}{cc}
\cosh\chi & \sinh\chi\\
\sinh\chi & \cosh\chi
\end{array}\!\right)\negmedspace\left(\!\begin{array}{c}
\phi_{+}\\
\phi_{-}
\end{array}\!\right)\!,\,\tanh2\chi\negmedspace=\negmedspace-\frac{v_{0,0}}{v_{0}}.
\end{equation}
which yields a diagonal $V$-matrix with equal velocities $\sqrt{v_{0}^{2}-v_{0,0}^{2}}$
for the modes $\phi_{\pm}$. In the new basis, the pairing operator
is 
\begin{equation}
O_{p}=e^{i\sqrt{2}[\phi_{+,0}-\phi_{-,0}]}=e^{i\sqrt{2}(\cosh\chi-\sinh\chi)[\phi_{+}-\phi_{-}]}\,,
\end{equation}
and its scaling dimension is $\delta=2(\cosh\chi-\sinh\chi)^{2}$. 
In the limit of weak interaction, $\chi\approx-v_{0,0}/(2v_{0})$
and $\delta\approx2(1+\frac{v_{0,0}}{v_{0}})$. The pairing is relevant,
$\delta<2$, when $v_{0,0}<0$ (attractive interaction). 
The backscattering operator $O_{b}=e^{i\sqrt{2}[\phi_{+,0}+\phi_{-,0}]}$ has $\delta  = 2(\cosh\chi + \sinh\chi)^{2}$ and is irrelevant when  $v_{0,0}<0$. 

Ignoring the renormalization of $v_{0,0}$, the RG equation for the
dimensionless pairing amplitude $\lambda_{p}$ is 
\begin{equation}
\frac{d}{dl}\lambda_{p}=(2-\delta)\lambda_{p}
\end{equation}
and thus $\lambda_{p}(l)=e^{(2-\delta)l}\lambda_{p}(0)$. The strong
coupling scale $l_{\Delta}$ is found from $\lambda_{p}(l_{\Delta})\sim1$. Writing $D=D_{KFP}e^{-l}$,
we find 
\begin{equation}
\Delta_{n}\sim D_{KFP}|\lambda_{p}(0)|^{-1/(\delta-2)}\sim D_{KFP}|\lambda_{p}(0)|^{-v_{0}/(2v_{0,0})}\sim D_{KFP}|\lambda_{p}(0)|^{1/|\lambda_{0}^{z}|}
\end{equation}
where we introduced $\lambda_{0}^{z}=2v_{0,0}/v_{0}$ and took $v_{0,0}<0$
and $v_{0,0}/v_{0}\ll1$. 

\subsection{Solution of the RG equations (\ref{eq:Rg1})--(\ref{Rg2})} 
In this Section, we give details on how to solve Eqs.~(\ref{eq:Rg1})--(\ref{Rg2}) of the main text. For completeness, we replicate the equations below: 
\begin{equation}
	\frac{d}{dl}\lambda^{x}  =\lambda^{y}\lambda^{z}\,,\quad\frac{d}{dl}\lambda^{y}=\lambda^{x}\lambda^{z}\,,\quad
	\frac{d}{dl}\lambda^{z} =\lambda^{x}\lambda^{y}\,,\quad(l=\ln D_{\text{KFP}}/D)\,, 
\end{equation}
where $D\ll D_{\text{KFP}}$ is the reduced bandwidth and we take the initial condition $\boldsymbol{\lambda}(D_{\text{KFP}})=(\lambda_{p}+\lambda_{b},\lambda_{b}-\lambda_{p},\lambda_{0}^{z})^{T}$.

The above equation [Eqs.~(\ref{eq:Rg1})--(\ref{Rg2}) of the main text] can be solved  after identifying
	the two integrals of motion, $(\lambda^{z})^{2}-(\lambda^{x})^{2}=c_{x}$
	and $(\lambda^{z})^{2}-(\lambda^{y})^{2}=c_{y}$, with $c_{x,y}$
	constants. For illustration, we provide the solution in two limits
	$|\lambda_{p}|\ll|\lambda_{0}^{z}|$ and $|\lambda_{p}|\gg|\lambda_{0}^{z}|$.
	To illustrate the first limit, we take initial condition $-\lambda_{0}^{z}\gg\lambda^{x}=-\lambda^{y}=\lambda_{p}>0$.
	One then finds $\lambda^{x}(l)=-\lambda^{y}(l)=-\sqrt{c_{x}}\text{csch}\left(l\sqrt{c_{x}}-\sinh^{-1}\frac{\sqrt{c_{x}}}{\lambda_{p}}\right)$
	and $\lambda^{z}(l)=\sqrt{c_{x}}\coth\left(l\sqrt{c_{x}}-\sinh^{-1}\frac{\sqrt{c_{x}}}{\lambda_{p}}\right)$.
	We can define the strong coupling limit as $|\lambda^{x}(l_{\Delta})|\approx|\lambda^{z}(l_{\Delta})|$
	which yields $l_{\Delta}\approx\frac{\ln2\frac{|\lambda_{0}^{z}|}{\lambda_{p}}}{|\lambda_{0}^{z}|}$
	in the perturbative regime, $\lambda_{p}\ll|\lambda_{0}^{z}|\ll1$.
	From this, we obtain the scale $\Delta_{n} = D_{\text{KFP}}e^{-l_{\Delta}} \approx  D_{\text{KFP}}(2\frac{|\lambda_{0}^{z}|}{\lambda_{p}})^{-\frac{1}{|\lambda_{0}^{z}|}}$.
	In the limit $|\lambda_{p}|\gg|\lambda_{0}^{z}|$, we find $\lambda^{x}(l)=-\lambda^{y}(l)=\lambda_{p}\sec(l|\lambda_{p}|)$
	and $\lambda^{z}(l)=-|\lambda_{p}|\tan(l|\lambda_{p}|)$. Solving
	the strong-coupling condition yields $l_{\Delta}\approx\frac{\pi}{2|\lambda_{p}|}$
	in the perturbative regime, $|\lambda_{p}|\ll1$. From this, we obtain
	the scale $\Delta_{n}\approx D_{\text{KFP}}e^{-\pi/2|\lambda_{p}|}$.
	In the limit $|\lambda_{p}|\sim|\lambda_{0}^{z}|$, the two expressions
	for $\Delta_{n}$ approximately agree, $\Delta_{n}\sim D_{\text{KFP}}e^{-c/|\lambda_{p}|}$
	with $c\sim1$. 

\subsection{Orthogonal transformation $S$}

In this Section, we give more details on the transformation, Eq.~(\ref{eq:JTransformation})
of the main text, that is used to gauge out the disorder term of the
neutralons. 

The orthogonal matrix $S_{\tau}$ is found by requiring that the linear-in-$\mathbf{\tilde{J}}_{\tau}$
terms cancel in the Hamiltonian $H_{\text{neutral}}+H_{\text{pairing}}$,
see Eqs.~(\ref{eq:H_pairing1}), (\ref{eq:HNeutral}). One finds
that $S_{\tau}$ satisfies the condition 
\begin{equation}
\frac{1}{8\pi}\sum_{jk}\varepsilon^{ijk}[S_{\tau}\partial_{x}S_{\tau}^{T}]^{jk}=v_{0}^{-1}\frac{\frac{\lambda^{i}}{2\pi}\xi_{\overline{\tau}}^{i}(x)-\frac{2}{3}\xi_{\tau}^{i}(x)}{(\frac{2}{3})^{2}-(\frac{1}{2\pi}\lambda^{i})^{2}}\,,\quad(i=x,y,z)
\end{equation}
where $\xi_{\tau}^{z}=0$, $\xi_{\tau}^{x}=\xi_{\tau}+\xi_{\tau}^{*}$,
and $\xi_{\tau}^{y}=i(\xi_{\tau}-\xi_{\tau}^{*})$. The matrix $S_{\tau}\partial_{x}S_{\tau}^{T}=-(\partial_{x}S_{\tau})S_{\tau}^{T}$
is antisymmetric, and we find 
\begin{equation}
\partial_{x}S_{\tau}=-U_{\tau}S_{\tau}\,,\quad U_{\tau}^{lm}(x)\equiv-\sum_{i}\varepsilon^{ilm}4\pi v_{0}^{-1}\frac{\frac{2}{3}\xi_{\tau}^{i}(x)-\frac{1}{2\pi}\lambda^{i}\xi_{\overline{\tau}}^{i}(x)}{(\frac{2}{3})^{2}-(\frac{1}{2\pi}\lambda^{i})^{2}}\,.
\end{equation}
The solution $S_{\tau}(x)$ can be written as a path-ordered exponential,
\begin{equation}
S_{\tau}(x)=T_{x}\exp\left[\int_{x_{0}}^{x}dx'U_{\tau}(x')\right]S_{\tau}(x_{0})\,,\label{eq:S}
\end{equation}
and $x=x_{0}$ denotes the starting point of the disordered region;
we will thus take $S_{\tau}(x_{0})=1$. 

Let us next focus on the realistic case where disorder couples to
the adjacent top and bottom layers in equal strengths, $\xi_{t}=\xi_{b}^{*}$.
We then have $\xi_{t}^{x}=\xi_{b}^{x}$ and $\xi_{t}^{y}=-\xi_{b}^{y}$
and 
\begin{equation}
U_{\tau}(x)=4\pi v_{0}^{-1}\left(\begin{array}{ccc}
0 & 0 & \frac{1}{\frac{2}{3}-\frac{1}{2\pi}\lambda^{y}}\xi_{\tau}^{y}(x)\\
0 & 0 & -\frac{1}{\frac{2}{3}+\frac{1}{2\pi}\lambda^{x}}\xi_{\tau}^{x}(x)\\
-\frac{1}{\frac{2}{3}-\frac{1}{2\pi}\lambda^{y}}\xi_{\tau}^{y}(x) & \frac{1}{\frac{2}{3}+\frac{1}{2\pi}\lambda^{x}}\xi_{\tau}^{x}(x) & 0
\end{array}\right)\,.
\end{equation}
 Due to the layer-correlated disorder, we have the non-trivial
property $S_{\tau}(x)^{T}\text{diag}(-1,1,1)S_{\overline{\tau}}(x)=\text{diag}(-1,1,1)$.
In particular, the pairing term $\mathbf{J}_{t}^{T}\text{diag}(-\lambda,\lambda,\lambda)\mathbf{J}_{b}$
is invariant under the transformation~(\ref{eq:JTransformation})
of the main text. 

We still cannot evaluate $S(x)$ in explicit form, but we can obtain
its average properties. The matrix $S(x)$ describes a sequence (on
the $x$-axis) of random rotations by a random angle $\sqrt{\xi_{x}^{2}+\xi_{y}^{2}}$
about a random axis $(\xi_{x},\xi_{y},0)$. We will next assume that
the correlation length of the angle distribution is much shorter than
that of the axis direction. In this limit we can carry out the average
in two steps, first over the angle and then over the axis direction.
For a fixed axis, the path-ordering in Eq.~(\ref{eq:S}) can be removed
and $S_{\tau}(x)$ can be obtained explicitly. For example, for a
rotation about the $x$-axis, we can set $\xi_{y}(x)=0$ and find
\begin{equation}
S_{t}(x)=S_{b}(x)=\left(\begin{array}{ccc}
1 & 0 & 0\\
0 & \cos\tilde{v}_{0}^{-1}\int_{x_{0}}^{x}dx'\xi_{x}(x') & -\sin\tilde{v}_{0}^{-1}\int_{x_{0}}^{x}dx'\xi_{x}(x')\\
0 & \sin\tilde{v}_{0}^{-1}\int_{x_{0}}^{x}dx'\xi_{x}(x') & \cos\tilde{v}_{0}^{-1}\int_{x_{0}}^{x}dx'\xi_{x}(x')
\end{array}\right)\,.\label{eq:SM:S-x-axis}
\end{equation}
We denote here $\tilde{v}_{0}^{-1}=v_{0}^{-1}\frac{4\pi}{\frac{2}{3}+\frac{1}{2\pi}\lambda^{x}}$.
Averaging over the angle with $\left\langle \xi^{x}(x)\xi^{x}(x')\right\rangle =2a^{-1}W\delta(x-x')$
yields (we denote $S\equiv S_{t}=S_{b}$)
\begin{equation}
\left\langle S(x)\right\rangle =\left(\begin{array}{ccc}
1 & 0 & 0\\
0 & e^{-\tilde{v}_{0}^{-2}a^{-1}W(x-x_{0})} & 0\\
0 & 0 & e^{-\tilde{v}_{0}^{-2}a^{-1}W(x-x_{0})}
\end{array}\right)\,.
\end{equation}
Taking a Gaussian distribution for $\xi_{x}$ in Eq.~(\ref{eq:SM:S-x-axis}),
we can similarly find the higher moments such as $\left\langle S^{ij}(x)S^{kl}(x)\right\rangle $. 

After averaging over the angle, we can average over the axis vector
in the $x$-$y$ plane, which yields 
\begin{equation}
\overline{\left\langle S(x)\right\rangle }=\left(\begin{array}{ccc}
\frac{1}{2}(1+e^{-\tilde{v}_{0}^{-2}a^{-1}W(x-x_{0})}) & 0 & 0\\
0 & \frac{1}{2}(1+e^{-\tilde{v}_{0}^{-2}a^{-1}W(x-x_{0})}) & 0\\
0 & 0 & e^{-\tilde{v}_{0}^{-2}a^{-1}W(x-x_{0})}
\end{array}\right)\,.
\end{equation}
In the limit $(x-x_{0})\gg a\tilde{v}_{0}^{2}/W$, we find then $\overline{\left\langle S(x)\right\rangle }=\text{diag}(\frac{1}{2},\frac{1}{2},0)$.
For the matrix $M^{ij}(x)\equiv S_{t}^{zi}(x)S_{b}^{zj}(x)$ {[}see
Eq.~(\ref{eq:PairingTransformation}) of the main text{]}, we find
\begin{equation}
M(x)=\left(\begin{array}{ccc}
0 & 0 & 0\\
0 & \sin^{2}\tilde{v}_{0}^{-1}\int_{x_{0}}^{x}dx'\xi_{x}(x') & \frac{1}{2}\sin2\tilde{v}_{0}^{-1}\int_{x_{0}}^{x}dx'\xi_{x}(x')\\
0 & \frac{1}{2}\sin2\tilde{v}_{0}^{-1}\int_{x_{0}}^{x}dx'\xi_{x}(x') & \cos^{2}\tilde{v}_{0}^{-1}\int_{x_{0}}^{x}dx'\xi_{x}(x')
\end{array}\right)\,,
\end{equation}
and $\left\langle M\right\rangle =\frac{1}{2}\text{diag}(0,1-e^{-4\tilde{v}_{0}^{-2}a^{-1}W(x-x_{0})},1+e^{-4\tilde{v}_{0}^{-2}a^{-1}W(x-x_{0})})$.
For the variance of $M(x')$, we find {[}in the limit $x^{(\prime)}-x_{0}\gg a\tilde{v}_{0}^{2}/W${]}
\begin{equation}
\left\langle M(x)M(x^{\prime})\right\rangle -\left\langle M\right\rangle ^{2}=\begin{cases}
0\,, & x-x^{\prime}\gg a\tilde{v}_{0}^{2}/W\,,\\
\left\langle M\right\rangle \,, & x-x^{\prime}\ll a\tilde{v}_{0}^{2}/W\,.
\end{cases}
\end{equation}
Upon averaging over the axis and taking the limit $(x-x_{0})\gg a\tilde{v}_{0}^{2}/W$,
we find 
\begin{flalign}
\overline{\left\langle M\right\rangle } & =\frac{1}{4}\left(\begin{array}{ccc}
1 & 0 & 0\\
0 & 1 & 0\\
0 & 0 & 2
\end{array}\right)\,. \label{eq:SMO2}
\end{flalign}
This form was used to obtain the average of $\mathcal{M}=M-\text{diag}(0,0,1)$
in the main text, Eq.~(\ref{eq:HneutralPairingRotated}). 

Even though this result was obtained in the limit where the angle
$\sqrt{\xi_{x}^{2}+\xi_{y}^{2}}$ varies much faster than the axis
direction, we expect the result to hold more generally since for random
disorder we sample all points on the Bloch sphere in an uncorrelated way.

\subsection{The fixed point $\boldsymbol{\lambda}=(\lambda,-\lambda,\lambda)^{T}$ with $\lambda < 0$ in the disordered case}

The other strong-pairing fixed point discussed below Eqs.~(\ref{eq:Rg1})--(\ref{Rg2}) of the main text 
has $\boldsymbol{\lambda}=(\lambda,-\lambda,\lambda)^{T}$ with $\lambda<0$.
This vector is not invariant under the gauge transformation Eq.~(\ref{eq:JTransformation}) of the main text.
We have instead 
\begin{equation}
	\mathbf{J}_{t}^{T}\boldsymbol{\tau}\mathbf{J}_{b}=\mathbf{\tilde{J}}_{t}^{T}\boldsymbol{\tau}\mathbf{\tilde{J}}_{b}+2\mathbf{\tilde{J}}_{t}^{T}\boldsymbol{\mathcal{M}}\mathbf{\tilde{J}}_{b}\,,
\end{equation}
where $\boldsymbol{\tau}=\text{diag}(1,-1,1)$ and  $\mathcal{M}^{ij}(x)=S_{t}^{zi}(x)S_{b}^{zj}(x)-\delta^{ij}\delta^{iz}$
depends on position. However, for a suitable model of disorder, the
matrix $\mathcal{M}(x)$ can be separated into a random and a uniform
(position-independent) part. The random part is RG irrelevant 
and we can neglect it at low energies. The uniform part is $O(2)$
symmetric, Eq.~(\ref{eq:SMO2}), $\mathcal{M}=\text{diag}(\frac{1}{4},\frac{1}{4},-\frac{1}{2})$.
Thus, upon averaging over disorder, we obtain a non-disordered Hamiltonian of the
form Eq.~(\ref{eq:HneutralPairingRotated}) of the main text, with $\boldsymbol{\lambda}=\lambda(\frac{3}{2},-\frac{1}{2},0)^{T}$.
We can then use the Eqs.~(\ref{eq:Rg1})--(\ref{Rg2}) of the main text to study
the renormalization group flow. Now $\lambda^{z}=0$ initially, but
flows to negative values since $\lambda^{x}\lambda^{y}<0$. We therefore
expect to reach the strong-pairing fixed point $\boldsymbol{\lambda}=(\lambda,-\lambda,\lambda)^{T}$
with $\lambda<0$.

\subsection{Four-neutralon pairing and neutralon sectors}

\begin{figure}
\includegraphics[width=0.4\columnwidth]{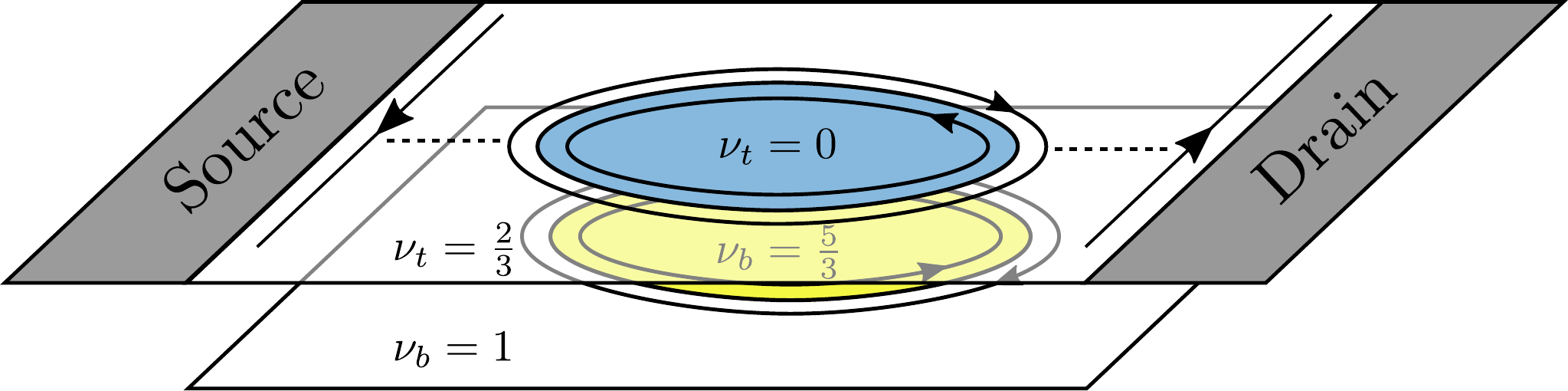}\quad\includegraphics[width=0.45\columnwidth]{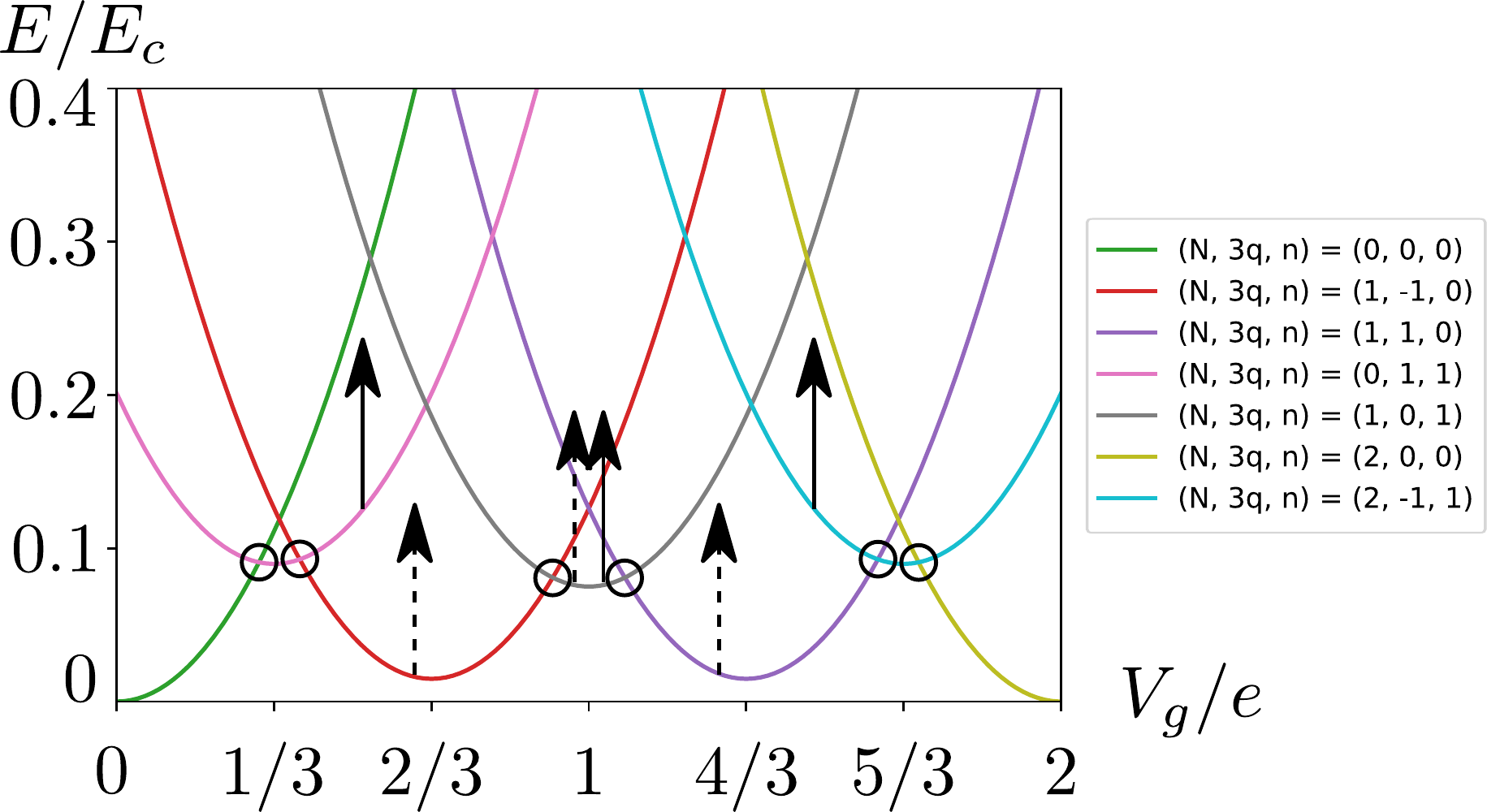}

\caption{Left: A Coulomb blockaded antidot in the top layer can be used to
access a signature of the neutralon pairing. In charge transport from
left to right, electrons or quasiparticles can enter the antidot.
The current is highest near charge degeneracy points (crossings of
Coulomb parabolas) shown on the right. Right: Spectrum of the ``charging
energy'' Hamiltonian from Eq.~(\ref{eq:Hc}) as a function of gate
charge. Charge degeneracy points of the ground state manifold are
shown in circles. The solid/dashed arrows indicate those states whose
energies would be lifted up by non-zero neutralon/electron pairing.
Neutralon pairing is therefore distinct from electron pairing. In
the limit of a large antidot (large circumference $L$), the charging
energies scale as $E_{c}\propto1/L$. In that limit the neutralon
pairing gap is large compared to $E_{c}$. The parameters used are
$E_{cq}=0.133E_{c}$, $E_{cn}=0.075E_{c}$. \label{fig:QDs}}
\end{figure}

In this section we show that the neutralon pairing is between 4 neutralons.
We also propose a conceptual setup that facilitates the measurement
of 4-neutralon pairing.

On a single edge of the bilayer system, say the top edge in Fig.~\ref{fig:QDs},
the edge Fock space consists of sectors differing by their numbers
of neutralons, quasiparticles, and electrons. Let us label the different
sectors by $(n_{e},n_{e/3},n_{0})$ where $n_{e}$ is the number of
electrons on the edge added from outside, $n_{e/3}$ is the number
of number of fractional $e/3$ quasiparticles added to the edge from
the strongly-correlated bulk, and $n_{0}$ is the number of neutralons
on the edge~\citep{PhysRevLett.114.156401}. The creation operator
of a single neutralon is $e^{\mp i\phi_{\pm,0}/\sqrt{2}}$ (where
$\pm$ is the direction of propagation, or layer index). However,
the physical operators that can appear in the Hamiltonian are combinations
of electron or quasiparticle creation operators. The operators that
create an electron on the, say, top edge are given by $e^{i\phi_{-1}}=e^{-i\sqrt{\frac{1}{2}}\phi_{+0}}e^{i\sqrt{\frac{3}{2}}\phi_{-2/3}}$
and $e^{2i\phi_{-1}}e^{i3\phi_{1/3}}=e^{i\sqrt{\frac{1}{2}}\phi_{+0}}e^{i\sqrt{\frac{3}{2}}\phi_{-2/3}}$
and both change the neutralon number by one. The neutral combination
of these operators is $e^{-i\phi_{-1}}e^{-i3\phi_{1/3}}=e^{-i\sqrt{2}\phi_{+0}}$
which creates a pair of neutralons. This term appears in the KFP~\citep{kane_randomness_1994}
action, Eq.~(\ref{eq:SKFPFP}). The pairing term in Eq.~~(\ref{eq:H_pairing1})
is $e^{-i\sqrt{2}[\phi_{+0}-\phi_{-0}]}$ and creates 2 neutralons
to each edge. Thus, we call it 4-neutralon pairing. 

We also note that a charge-$2e$ operator $e^{3i\phi_{-1}}e^{i3\phi_{1/3}}=e^{2i\sqrt{\frac{3}{2}}\phi_{-2/3}}$,
does not change the neutralon number.   Finally, we have the operators
$e^{-i\phi_{1/3}}=e^{-i\sqrt{\frac{1}{2}}\phi_{+0}}e^{i\sqrt{\frac{1}{6}}\phi_{-2/3}}$
and $e^{i\phi_{-1}}e^{2i\phi_{1/3}}=e^{i\sqrt{\frac{1}{2}}\phi_{+0}}e^{i\sqrt{\frac{1}{6}}\phi_{-2/3}}$ that
add an $e/3$ quasiparticle and add/remove a neutralon. Thus, starting
from a reference sector, say $(0,0,0)$, we can access the sectors
$(n_{e},n_{e/3},n_{e}+n_{e/3}+2m)$, where $n_{e/3},n_{e},m$ are
integers. 

By using quantum dots or antidots the different neutralon sectors
can be in principle accessed, see Fig.~\ref{fig:QDs}. The charge
states of an antidot in one layer can be labeled by $(N,q,n)$, where
$N$ is the number of electrons, $q$ is the number of quasiparticles,
and $n$ is the number of neutralons. For the antidot, we take the
charging energy Hamiltonian of Ref.~\citep{PhysRevLett.114.156401}:
\begin{equation}
H_{c}=E_{c}(N+3q-\frac{1}{e}V_{g})^{2}+E_{cq}q^{2}+E_{cn}n^{2}\,.\label{eq:Hc}
\end{equation}
Here $E_{c}$ is the total charging energy, coupling to electrons
and quasiparticles, and $E_{cq}$ and $E_{cn}$ are separate ``charging
energies'' for quasiparticles and neutralons. The spectrum from Eq.~(\ref{eq:Hc})
is plotted in Fig.~\ref{fig:QDs}b. 

\subsection{Mode expansion and the four neutralon parity sectors }

In this section, we show that there are two degenerate ground states
for the neutralon pairing operator. These ground states correspond
to two of the four ``parity'' sectors, defined by the neutralon number modulo
4. In a finite-size edge, the degeneracy between the sectors is split
by the Hamiltonian Eq.~(\ref{eq:Hc}). 

We introduce the mode expansion~\citep{2017AnPhy.385..287P} for
neutralon $\phi_{\pm,0}$ in a periodic edge of length $L$: 
\begin{equation}
\phi_{\tau,0}(x)=\frac{2\pi x}{L\sqrt{2}}N_{\tau,0}-\tau\sqrt{2}\chi_{\tau,0}-i\sum_{q=2\pi m/L>0}\sqrt{\frac{2\pi}{Lq}}[e^{\tau iqx}b_{q,\tau}-e^{-\tau iqx}b_{q,\tau}^{\dagger}]\,,\label{eq:ModeExp}
\end{equation}
where $m$ is a positive integer, $[b_{q},b_{q'}^{\dagger}]=\delta_{q,q'}$
and $[\chi_{\tau,0},N_{\tau',0}]=i\delta_{\tau'\tau}$. We can check
that 
\begin{equation}
[\phi_{\tau,0}(x),\phi_{\tau,0}(x')]=\tau i\pi\text{sgn}(x-x')\,,
\end{equation}
by using the identity (as $\alpha\to0$)
\begin{equation}
\sum_{n=1}^{\infty}\frac{1}{n}e^{-\alpha n}\sin\frac{2\pi n}{L}(x-x')=\frac{\pi}{2}\text{sgn}(x-x')-\frac{\pi}{L}(x-x')\,.
\end{equation}

We will consider a homogeneous neutralon pairing operator $\cos\sqrt{2}[\phi_{+,0}-\phi_{-,0}]$.
We will look for a homogeneous configuration: the pairing energy is
minimized when the operator $\chi_{+,0}+\chi_{-,0}$ is pinned to
a value $(n+\frac{1}{2})\pi$. Let us find which of the $\chi_{+,0}+\chi_{-,0}$-eigenstates
$|(n+\frac{1}{2})\pi\rangle$ can be called equivalent. For this,
we note that the operator conjugate to $\chi_{+,0}+\chi_{-,0}$ is
$(N_{+,0}+N_{-,0})/2$, i.e., $[\chi_{+,0}+\chi_{-,0},\frac{1}{2}(N_{+,0}+N_{-,0})]=i$.
In the number-basis, the latter operator takes half-integer values.
The operator $e^{i(\chi_{+,0}+\chi_{-,0})}$ is the raising operator
in the number basis: $e^{i(\chi_{+,0}+\chi_{-,0})}=\sum_{n\in\mathbb{Z}}|\frac{1}{2}n+1\rangle\langle\frac{1}{2}n|$
as follows from the commutation relation $[e^{i(\chi_{L}+\chi_{R})},\frac{1}{2}(N_{+,0}+N_{-,0})]=-e^{i(\chi_{L}+\chi_{R})}$.
In the number basis, the eigenstates of $e^{i(\chi_{+,0}+\chi_{-,0})}$
are thus given by 
\begin{equation}
|k\rangle=\sum_{n\in\mathbb{Z}}e^{-\frac{1}{2}ink}|\frac{1}{2}n\rangle\,,
\end{equation}
with eigenvalue $e^{ik}$. We note that $|k\rangle$ and $|k+4\pi\rangle$
are the same state. Therefore, we can define $k$ in a ``Brillouin
zone'' $k\in[0,4\pi)$. Thus, we have four inequivalent eigenstates
of the homogeneous pairing Hamiltonian: $|\frac{1}{2}\pi\rangle$,
$|\frac{3}{2}\pi\rangle$, $|\frac{5}{2}\pi\rangle$, and $|\frac{7}{2}\pi\rangle$.
 Out of these states, we can construct eigenstates of neutralon number
parity mod 4, $e^{i\pi\frac{1}{2}(N_{+,0}+N_{-,0})}$ . In order to
do this, we note that $e^{i\pi\frac{1}{2}(N_{+,0}+N_{-,0})}|k\rangle=|k-\pi\rangle$.
We find, 
\begin{flalign}
|0\rangle & =|\frac{1}{2}\pi\rangle+|\frac{5}{2}\pi\rangle+|\frac{3}{2}\pi\rangle+|\frac{7}{2}\pi\rangle\,,\label{eq:paritystate0}\\
|1\rangle & =|\frac{1}{2}\pi\rangle-|\frac{5}{2}\pi\rangle-i|\frac{3}{2}\pi\rangle+i|\frac{7}{2}\pi\rangle\,,\\
|2\rangle & =|\frac{1}{2}\pi\rangle+|\frac{5}{2}\pi\rangle-|\frac{3}{2}\pi\rangle-|\frac{7}{2}\pi\rangle\,,\\
|3\rangle & =|\frac{1}{2}\pi\rangle-|\frac{5}{2}\pi\rangle+i|\frac{3}{2}\pi\rangle-i|\frac{7}{2}\pi\rangle\,.\label{eq:paritystate3}
\end{flalign}
The state $|n\rangle$ has an eigenvalue $e^{i\pi n/2}$ of ``parity''
mod 4, $e^{i\pi\frac{1}{2}(N_{+,0}+N_{-,0})}$. 

However, there is an additional restriction on the states (\ref{eq:paritystate0})--(\ref{eq:paritystate3}).
In order to have a homogeneous solution we require that $(\phi_{+,0}-\phi_{-,0})|_{x}^{x+L}=0$,
or $N_{+,0}=N_{-,0}$~\citep{PhysRevB.84.195436}. Considering states
where $N_{\pm,0}$ are integers, we see that the neutralon number
parity is $\pm1$. Thus, only the states $|0\rangle$ and $|2\rangle$
are allowed ground states. For these two states, we have the boundary
conditions for the neutralon creation operator $e^{\mp i\phi_{\pm,0}(x)/\sqrt{2}}$
as $e^{\mp i\phi_{\pm,0}(x+L)/\sqrt{2}}=e^{\mp i\pi n/2}e^{\mp i[\phi_{\pm,0}(x)/\sqrt{2}]}$
where $n=0,2$.  Thus, each ground state corresponds to a unique
boundary condition for the neutralon creation operator. Upon changing
$N_{\tau,0}\to N_{\tau,0}+1$, the operator $e^{\tau i\phi_{\tau,0}(L)/\sqrt{2}}$
acquires a minus sign, which can be interpreted as a braiding phase
$\pi$, resulting from the statistical angle $\pi/2$ for neutralons
(semions)~\cite{PhysRevB.36.4581}. 

Adding/removing a single neutralon on/from one of the edges will
violate the condition $N_{+,0}=N_{-,0}$ (and changes the state as
$|n\rangle\to|n\pm1\rangle$) and will therefore come at a cost $\Delta_{n}$
(the pairing energy). This energy cost can be used as a signature
of neutralon pairing, as is illustrated in Fig.~\ref{fig:QDs}b.

\subsection{Alternative QPC designs}

We show here additional QPC designs to Fig.~\ref{fig:QPC} of the
main text. In Fig.~\ref{fig:QPC_alt} we show two alternative designs.
Both designs are similar to that of Fig.~\ref{fig:QPC}b in that
only electrons are allowed to tunnel through the middle section. Hence,
in particular, similar considerations to those made for Fig.~\ref{fig:QPC}b
show that the tunneling current across these alternative QPC designs
would be strongly suppressed as well. 
\begin{figure}[b]
\includegraphics[width=0.95\columnwidth]{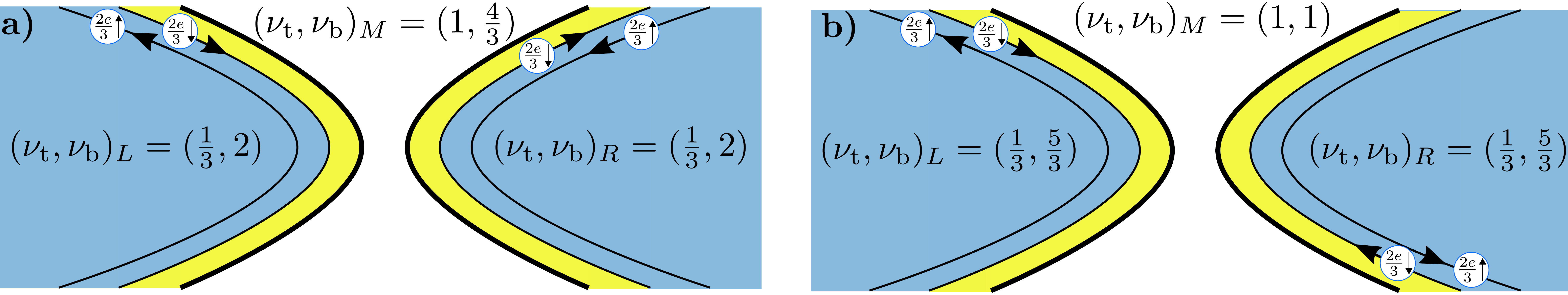}

\caption{Two alternative QPC designs. \label{fig:QPC_alt}}
\end{figure}

\end{widetext}
\end{document}